\RequirePackage{fix-cm}
\documentclass[twocolumn,epjc3]{svjour3}
\smartqed  
\RequirePackage{graphicx}
\RequirePackage{multirow}
\RequirePackage{amstext}
\RequirePackage{rotating}
\journalname{Eur. Phys. J. C}
\begin{document}

\title{Exotic tetraquark states with the $qq\bar{Q}\bar{Q}$ configuration
}

\author{Si-Qiang Luo\thanksref{addr1,addr2},
        Kan Chen\thanksref{addr1,addr2},
        Xiang Liu\thanksref{e1,addr1,addr2},
        Yan-Rui Liu\thanksref{e2,addr3},
        Shi-Lin Zhu\thanksref{e3,addr4,addr5,addr6}}
\thankstext{e1}{e-mail: xiangliu@lzu.edu.cn}
\thankstext{e2}{e-mail: yrliu@sdu.edu.cn}
\thankstext{e3}{e-mail: zhusl@pku.edu.cn}
\institute{School of Physical Science and Technology, Lanzhou University, Lanzhou 730000, China \label{addr1}\and
Research Center for Hadron and CSR Physics, Lanzhou University and Institute of Modern Physics of CAS, Lanzhou 730000, China \label{addr2}\and
School of Physics and Key Laboratory of Particle Physics and Particle Irradiation (MOE), Shandong University, Jinan 250100, China\label{addr3} \and
School of Physics and State Key Laboratory of Nuclear Physics and Technology, Peking University, Beijing 100871, China \label{addr4} \and
Collaborative Innovation Center of Quantum Matter, Beijing 100871, China \label{addr5} \and
Center of High Energy Physics, Peking University, Beijing 100871, China \label{addr6}
}

\date{Received: date / Accepted: date}

\maketitle

\begin{abstract}
In this work, we study systematically the mass splittings of the
$qq\bar{Q}\bar{Q}$ ($q=u$, $d$, $s$ and $Q=c$, $b$) tetraquark
states with the color-magnetic interaction by considering color
mixing effects and estimate roughly their masses. We find that the
color mixing effect is relatively important for the $J^P=0^+$ states
and possible stable tetraquarks exist in the $nn\bar{Q}\bar{Q}$
($n=u$, $d$) and $ns\bar{Q}\bar{Q}$ systems either with $J=0$ or
with $J=1$. Possible decay patterns of the tetraquarks are briefly
discussed.
\end{abstract}

\section{Introduction}\label{sec1}

Searching for exotic hadronic states is an interesting research
topic full of opportunities and challenges. In the past decade, the
reported charmonium-like states like $X(3872)$ \cite{Choi:2003ue},
$Y(3940)$ \cite{Abe:2004zs}, $Y(4140)$ \cite{Aaltonen:2009tz,Aaltonen:2011at},
$Z^+(4430)$ \cite{Choi:2007wga}, and $Z_c^+(4200)$
\cite{Chilikin:2014bkk}, the bottomonium-like states $Z_b(10610)$
and $Z_b(10650)$ \cite{Belle:2011aa}, and the open-heavy flavor
meson $X(5568)$ \cite{D0:2016mwd} have stimulated extensive
discussions on their exotic assignments. The interested readers may
refer to the recent literatures
\cite{Liu:2013waa,Chen:2016qju,Chen:2016spr,Lebed:2016hpi,Esposito:2016noz,Hosaka:2016pey,Richard:2016eis}
for the comprehensive review of progress.

Among the various exotic state assignments, the tetraquark
configuration is most popular in explaining these observed novel
phenomena. Thus, studying tetraquark states has become an important
issue of exploring exotic hadronic matter. Where and how to identify
tetraquark states from these observed $XYZ$ states is a main task.
Usually, it is easy to identify a hadronic state as an exotic one if
it has exotic quantum numbers like $0^{--}$, $0^{+-}$, $1^{-+}$,
$2^{+-}$, and so on. In order to identify a tetraquark state, we
need to pay attention to not only its quark component but also its
special properties. If a hadronic state has valence quarks with four
different flavors, we may conclude that this hadron is probably a
tetraquark state when the molecule interpretation is not favored.
The newly observed $X(5568)$ in the $B_s^0\pi^\pm$ channel
\cite{D0:2016mwd} is a typical example since the $X(5568)$ contains
valence quarks of four different flavors and the $\bar{B}K$
interaction in the isovector channel is not strong enough to form a
molecule. Besides, if a hadron has the $qq\bar{Q}\bar{Q}$
configuration, where $Q$ denotes $b$ or $c$ quark and $q$ is a light
quark, we may identify it to be a tetraquark state when it is far
below relevant meson-meson thresholds.

Until now, the tetraquark states with the $qq\bar{Q}\bar{Q}$
configuration have not been reported by experiments. On the other
hand, the existence and stability of such states have been discussed
by theorists for a long time. Different models
\cite{Ader:1981db,Ballot:1983iv,Lipkin:1986dw,Zouzou:1986qh,Heller:1986bt,Carlson:1987hh,Manohar:1992nd,SilvestreBrac:1993ry,SilvestreBrac:1993ss,Semay:1994ht,Pepin:1996id,Brink:1998as,SchaffnerBielich:1998ci,Vijande:2006jf,Cui:2006mp,Zhang:2007mu,Ebert:2007rn,Navarra:2007yw,Lee:2007tn,Yang:2009zzp,Vijande:2009kj,Carames:2011zz,Vijande:2013qr,Hyodo:2012pm,hyodoetal}
with various potentials as well as the associated interactions were introduced to describe the $qq\bar{Q}\bar{Q}$ system, suggesting that the $qq\bar{Q}\bar{Q}$ states are stable against breakup into the $q\bar{Q}-q\bar{Q}$ meson pair. In the framework of QCD sum rule, the mass spectrum of the $qq\bar{Q}\bar{Q}$ states has been studied in Refs. \cite{Du:2012wp,Chen:2013aba}. 
In addition, Lattice QCD simulations gave us more hints on this
issue
\cite{Stewart:1998hk,Michael:1999nq,Cook:2002am,Doi:2006kx,Detmold:2007wk,Wagner:2010ad,Bali:2010xa,Wagner:2011ev,Brown:2012tm,Ikeda:2013vwa,Wagenbach:2014oxa,Peters:2015tra,Peters:2016wjm,Peters:2016isf,Francis:2016hui}.
An intuitive picture \cite{Peters:2016wjm} in Lattice QCD is that in
a tetraquark with $qq\bar{Q}\bar{Q}$ configuration, when the two
heavy quarks are in a long separation, the gluon exchange force
between them is screened by the two light quarks. This system is
similar to a hydrogen molecule \cite{Du:2012wp}. Thus, a
$(q\bar{Q}-q\bar{Q})$ loosely bound state can be formed
\cite{Ohkoda:2012hv,Sakai:2017avl}. On the other hand, if the two
heavy quarks are in a short separation, the $\bar{Q}\bar{Q}$
component can form a color source. As a result, the exotic
tetraquark state $(\bar{Q}\bar{Q}-qq)$ can be generated after
pairing with the light component $qq$. Although many exotic
$qq\bar{Q}\bar{Q}$ tetraquarks were also obtained in Lattice QCD
simulation, the binding energy of such tetraquarks strongly depends
on the details of the dynamical models. To understand further the
properties of the tetraquark states with the $qq\bar{Q}\bar{Q}$
configuration, both theorists and experimentalists still need to
make more efforts to explore them.

In this work, we continue to pay attention to the tetraquark states
with the $qq\bar{Q}\bar{Q}$ configuration. We will again adopt the
framework of the simple color-magnetic interaction, although it is
not a dynamical model. Recently, we have systematically applied it
to tetraquark and pentaquark states in order to understand the
nature of the observed exotic hadrons and to predict exotic
phenomena \cite{Wu:2016vtq,Chen:2016ont,Wu:2016gas,Wu:2017weo}.
Here, we calculate the mass splittings of the tetraquark states with
the $qq\bar{Q}\bar{Q}$ configuration and estimate their mass
spectrum, with which we further discuss their decay patterns.
Hopefully, the information presented in this work will be helpful to
further experimental search for them.

This paper is organized as follows. After the introduction section,
we present the deduction of the elements of the interaction matrices
in Sec. \ref{sec2}. In Sec. \ref{sec3}, the adopted parameters and
numerical results are given in details. In the following Sec.
\ref{sec4}, we provide some discussions about these systems.
Finally, the paper ends with a short summary in Sec. \ref{sec5}.

\section{Formalism}\label{sec2}
The color magnetic interaction (CMI) in a hadron can be written as
\begin{eqnarray}
H_{CM}=-\sum\limits_{i<j}C_{ij}\lambda_i\cdot\lambda_j\sigma_i\cdot\sigma_j,
\end{eqnarray}
where the subscript $i$ denotes the $i$-th constituent quark in the
hadron and $\lambda_i$ and $\sigma_i$ represent the  Gell-Mann
matrices and the Pauli matrices, respectively. For an antiquark,
$\lambda_i$ should be replaced by $-\lambda_i^*$. This Hamiltonian
is deduced from part of the one gluon exchange interaction
\cite{DeRujula:1975qlm}. The $C_{ij}$ describes the effective
coupling constant between one quark and another quark or antiquark,
which incorporates the effects from the spatial wave function and
the quark mass. This constant will be estimated in the next section.

For the ground state of a tetraquark, the above CMI Hamitonian leads
to the mass formula
\begin{eqnarray}
H=\sum\limits_{i=1}^4 m_i+\langle H_{CM}\rangle
\end{eqnarray}
where $m_i$ is the effective mass of the $i$-th constituent quark.

We use the diquark-antidiquark bases to discuss the wave functions
of ground tetraquark states with the $qq\bar{Q}\bar{Q}$
configuration, where $Q=c$ or $b$ and $q=n$ or $s$ with $n=u$ or
$d$. Since we consider the mixing between different color-spin
structures, all the other bases will finally result in the same mass
splittings. In order to obtain all the ground states satisfying the
flavor-color-spin symmetry constraint from Pauli principle, we need
to exhaust all the possible spin and color wave functions of a
diquark-antidiquark system and combine them appropriately with the
flavor wave functions. With the obtained wave functions, one may
calculate the matrix elements using the approach illustrated in
Refs. \cite{Hogaasen:2004pm,Buccella:2006fn}.

The possible spin wave functions for the tetraquark states with the
$qq\bar{Q}\bar{Q}$ configuration are
\begin{eqnarray}
\chi_{1}=|(q_{1}q_{2})_{1}(\bar{Q}_{3}\bar{Q}_{4})_{1}\rangle_{2},\quad
\chi_{2}=|(q_{1}q_{2})_{1}(\bar{Q}_{3}\bar{Q}_{4})_{1}\rangle_{1},\nonumber\\
\chi_{3}=|(q_{1}q_{2})_{1}(\bar{Q}_{3}\bar{Q}_{4})_{1}\rangle_{0},\quad
\chi_{4}=|(q_{1}q_{2})_{1}(\bar{Q}_{3}\bar{Q}_{4})_{0}\rangle_{1},\\
\chi_{5}=|(q_{1}q_{2})_{0}(\bar{Q}_{3}\bar{Q}_{4})_{1}\rangle_{1},\quad
\chi_{6}=|(q_{1}q_{2})_{0}(\bar{Q}_{3}\bar{Q}_{4})_{0}\rangle_{0},\nonumber
\end{eqnarray}
where the subscripts denote the spins of the light diquark, the
heavy antidiquark, and the tetraquark state. The color wave function
can be analyzed by applying the SU(3) group theory, where the direct
product of the diquark-antidiquark components reads
\begin{eqnarray}
3_c\otimes 3_c\otimes \bar{3}_c\otimes \bar{3}_c=(6_c\oplus
\bar{3}_c)\otimes(\bar{6}_c\oplus3_c).
\end{eqnarray}
Obviously, we have two combinations to form a color singlet
tetraquark wave function, i.e.,
\begin{eqnarray}
\phi_{1}&=&|(q_{1}q_{2})^{6}(\bar{Q}_{3}\bar{Q}_{4})^{\bar{6}}\rangle\nonumber\\
        &=&\frac{1}{2\sqrt{6}}[2(rr\bar{r}\bar{r}+gg\bar{g}\bar{g}+bb\bar{b}\bar{b})+
rb\bar{b}\bar{r}+br\bar{b}\bar{r}\\
        &&+gr\bar{g}\bar{r}+rg\bar{g}\bar{r}+gb\bar{b}\bar{g}+bg\bar{b}\bar{g}+
gr\bar{r}\bar{g}+rg\bar{r}\bar{g}\nonumber\\
        &&+gb\bar{g}\bar{b}+bg\bar{g}\bar{b}+rb\bar{r}\bar{b}+br\bar{r}\bar{b}]\nonumber
\end{eqnarray}
and
\begin{eqnarray}
\phi_{2}&=&|(q_{1}q_{2})^{\bar{3}}(\bar{Q}_{3}\bar{Q}_{4})^{3}\rangle\nonumber\\
        &=&\frac{1}{2\sqrt{3}}(rb\bar{b}\bar{r}-br\bar{b}\bar{r}-gr\bar{g}\bar{r}+rg\bar{g}\bar{r}+
gb\bar{b}\bar{g}-bg\bar{b}\bar{g}\nonumber\\
        &&+gr\bar{r}\bar{g}-rg\bar{r}\bar{g}-gb\bar{g}\bar{b}+bg\bar{g}\bar{b}-rb\bar{r}\bar{b}+
br\bar{r}\bar{b}).
\end{eqnarray}

\begin{figure}[htbp]
\centering
\includegraphics[width=0.3\textwidth]{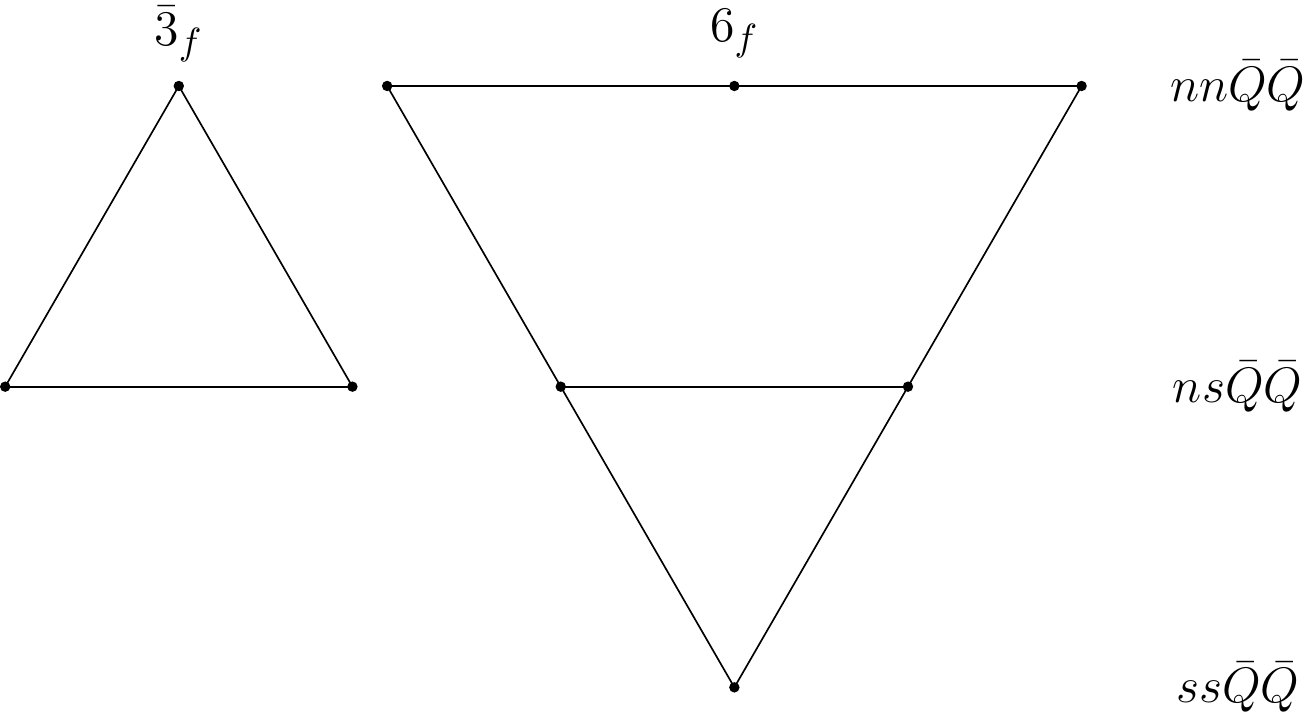}
\caption{Flavor representations for the $qq\bar{Q}\bar{Q}$ systems.
Here $Q=c,b$ and $q=n,s$ with $n=u,d$.}\label{flavorwave}
\end{figure}

In flavor space, the heavy quark is treated as an $SU(3)$ singlet
and we have states belonging to $6_f$ and $\bar{3}_f$ (Fig.
\ref{flavorwave}). For the $nn\bar{Q}\bar{Q}$ case, the isovector
states and the isoscalar states do not mix since we do not consider
isospin breaking effects. For the $ns\bar{Q}\bar{Q}$ case, the fact
$m_n\neq m_s$ leads to $SU(3)$ breaking and then the state mixing
between $6_f$ and $\bar{3}_f$.

Now we can combine the flavor, color, and spin wave functions
together. In this procedure, one needs to include the constraint
from the Pauli principle. In the diquark-antidiquark picture, the
possible wave function bases are
\begin{eqnarray}\label{basis-vectors}
&&\phi_1\chi_1=|(q_1q_2)^{6}_1(\bar{Q}_3\bar{Q}_4)^{\bar{6}}_1\rangle_2\delta_{12}^S\delta_{34},\nonumber\\
&&\phi_2\chi_1=|(q_1q_2)^{\bar{3}}_1(\bar{Q}_3\bar{Q}_4)^{3}_1\rangle_2\delta_{12}^A,\nonumber\\
&&\phi_1\chi_2=|(q_1q_2)^{6}_1(\bar{Q}_3\bar{Q}_4)^{\bar{6}}_1\rangle_1\delta_{12}^S\delta_{34},\nonumber\\
&&\phi_2\chi_2=|(q_1q_2)^{\bar{3}}_1(\bar{Q}_3\bar{Q}_4)^{3}_1\rangle_1\delta_{12}^A,\nonumber\\
&&\phi_1\chi_3=|(q_1q_2)^{6}_1(\bar{Q}_3\bar{Q}_4)^{\bar{6}}_1\rangle_0\delta_{12}^S\delta_{34},\nonumber\\
&&\phi_2\chi_3=|(q_1q_2)^{\bar{3}}_1(\bar{Q}_3\bar{Q}_4)^{3}_1\rangle_0\delta_{12}^A,\\
&&\phi_1\chi_4=|(q_1q_2)^{6}_1(\bar{Q}_3\bar{Q}_4)^{\bar{6}}_0\rangle_1\delta_{12}^S,\nonumber\\
&&\phi_2\chi_4=|(q_1q_2)^{\bar{3}}_1(\bar{Q}_3\bar{Q}_4)^{3}_0\rangle_1\delta_{12}^A\delta_{34},\nonumber\\
&&\phi_1\chi_5=|(q_1q_2)^{6}_0(\bar{Q}_3\bar{Q}_4)^{\bar{6}}_1\rangle_1\delta_{12}^A\delta_{34},\nonumber\\
&&\phi_2\chi_5=|(q_1q_2)^{\bar{3}}_0(\bar{Q}_3\bar{Q}_4)^{3}_1\rangle_1\delta_{12}^S,\nonumber\\
&&\phi_1\chi_6=|(q_1q_2)^{6}_0(\bar{Q}_3\bar{Q}_4)^{\bar{6}}_0\rangle_0\delta_{12}^A,\nonumber\\
&&\phi_2\chi_6=|(q_1q_2)^{\bar{3}}_0(\bar{Q}_3\bar{Q}_4)^{3}_0\rangle_0\delta_{12}^S\delta_{34}\nonumber.
\end{eqnarray}
Here, we present the wave functions with the notation
$|(q_1q_2)^{\rm color}_{\rm spin}(\bar{Q}_3\bar{Q}_4)^{\rm color }_{\rm spin }\rangle_{\text{total spin}}$. Since not all the wave
functions are allowed for a given set of quantum numbers, we have
introduced three factors $\delta_{12}^S$, $\delta_{12}^A$, and
$\delta_{34}$ to reflect this symmetry requirement. When the two
light quarks in flavor space are symmetric (antisymmetric), we have
$\delta_{12}^S=0$ ($\delta_{12}^A=0$). When the two heavy quarks are
identical, we have $\delta_{34}=0$. If the factor cannot be 0, its
value is set to be 1. We may easily ignore irrelevant wave functions
with these factors. Then the considered tetraquark states can be
categorized into six classes:
\begin{enumerate}\label{class}
\item The $(nn\bar{c}\bar{c})^{I=1}$, $(nn\bar{b}\bar{b})^{I=1}$, $ss\bar{c}\bar{c}$, and $ss\bar{b}\bar{b}$ states with $\delta_{12}^S=\delta_{34}=0$;
\item The $(nn\bar{c}\bar{c})^{I=0}$, $(nn\bar{b}\bar{b})^{I=0}$ states with $\delta_{12}^A=\delta_{34}=0$;
\item The $(nn\bar{c}\bar{b})^{I=1}$ and $ss\bar{c}\bar{b}$ states with $\delta_{12}^S=0$ and $\delta_{34}=1$;
\item The $(nn\bar{c}\bar{b})^{I=0}$ states with $\delta_{12}^A=0$ and $\delta_{34}=1$;
\item The $ns\bar{c}\bar{c}$ and $ns\bar{b}\bar{b}$ states with $\delta_{12}^S=\delta_{12}^A=1$ and $\delta_{34}=0$;
\item The $ns\bar{c}\bar{b}$ states with $\delta_{12}^S=\delta_{12}^A=\delta_{34}=1$.
\end{enumerate}

By selecting the corresponding tetraquark wave functions, we obtain
the CMI matrices for different cases. The expressions for classes 1
and 2, 3 and 4, and 5 and 6 are shown in tables \ref{class12},
\ref{class34}, and \ref{class56}, respectively. Here, we have
simplified the expressions by using the following definitions:
$\alpha=C_{12}+C_{34}$, $\beta =C_{13}+C_{14}+C_{23}+C_{24}$,
$\gamma=C_{13}+C_{14}-C_{23}-C_{24}$,
$\delta=C_{13}-C_{14}+C_{23}-C_{24}$, $\mu
=C_{13}-C_{14}-C_{23}+C_{24}$, $\eta  =C_{12}-3C_{34}$, and
$\theta=3C_{12}-C_{34}$.

\begin{table}[!h]
\caption{The quantum numbers, color-spin wave functions, and the
corresponding CMI matrices for the $nn\bar{c}\bar{c}$,
$nn\bar{b}\bar{b}$, $ss\bar{c}\bar{c}$, and $ss\bar{b}\bar{b}$
systems. Here, $(qq)^S$ ($(qq)^A$) means that the flavor wave
function of the two light quarks is symmetric
(antisymmetric).}\label{class12}
\begin{tabular}{cccc}\hline\hline
Symmetry&$J^P$&Wave functions&$\langle H_{CM}\rangle$\\\hline
$(qq)^S$&$0^+$&$(\phi_2\chi_3,\phi_1\chi_6)^T$&$\left(\begin{array}{cc}\frac{8}{3}(\alpha-\beta)  &2\sqrt{6}\beta\\&4\alpha\end{array}\right)$\\
&$1^+$&$(\phi_2\chi_2)$&$\frac{4}{3}(2\alpha-\beta)$\\
&$2^+$&$(\phi_2\chi_1)$&$\frac{4}{3}(2\alpha+\beta)$
\\\\
$(qq)^A$&$1^+$&$(\phi_2\chi_5, \phi_1\chi_4)^T$&
$\left(\begin{array}{cc}
-\frac{8}{3}\theta  &-2\sqrt{2}\beta  \\
                    &-\frac{4}{3}\eta
\end{array}\right)$\\\hline\hline
\end{tabular}
\end{table}

\begin{table*}[htbp]\centering
\caption{The quantum numbers, color-spin wave functions, and the
corresponding CMI matrices for the $nn\bar{c}\bar{b}$ and
$ss\bar{c}\bar{b}$ systems. Here, $(qq)^S$ ($(qq)^A$) means that the
flavor wave function of the two light quarks is symmetric
(antisymmetric).}\label{class34}
\begin{tabular}{cccc}\hline\hline
Symmetry&$J^P$&Wave functions&$\langle H_{CM}\rangle$\\\hline
$(qq)^S$&$0^+$&$(\phi_2\chi_3,\phi_1\chi_6)^T$
&$\left(\begin{array}{cc}
\frac{8}{3}(\alpha-\beta)  &2\sqrt{6}\beta\\
                           &4\alpha
\end{array}\right)$\\

&$1^+$&$(\phi_2\chi_4,\phi_2\chi_2,\phi_1\chi_5)^T$&
$\left(\begin{array}{ccc}
\frac{8}{3}\eta  &\frac{4\sqrt{2}}{3}\delta   &-2\sqrt{2}\beta\\
                 &\frac{4}{3}(2\alpha-\beta)  &4\delta\\
                 &                            &\frac{4}{3}\theta
\end{array}\right)$\\

&$2^+$&$(\phi_2\chi_1)$&$\frac{4}{3}(2\alpha+\beta)$\\\\

$(qq)^A$&$0^+$&$(\phi_2\chi_6, \phi_1\chi_3)^T$&
$\left(\begin{array}{cc}
-8\alpha    &2\sqrt{6}\beta                 \\
            &-\frac{4}{3}(\alpha+5\beta)
\end{array}\right)$\\

&$1^+$&$(\phi_2\chi_5, \phi_1\chi_4, \phi_1\chi_2)^T$&
$\left(\begin{array}{ccc}
-\frac{8}{3}\theta &-2\sqrt{2}\beta  &4\delta                       \\
                   &-\frac{4}{3}\eta &\frac{10\sqrt{2}}{3}\delta    \\
                   &                 &-\frac{2}{3}(2\alpha+5\beta)
\end{array}\right)$\\

&$2^+$&$(\phi_1\chi_1)$& $\frac{2}{3}(-2\alpha+5\beta)$
\\\hline\hline
\end{tabular}
\end{table*}

\begin{table*}[htbp]\centering
\caption{The quantum numbers, color-spin wave functions, and the
corresponding CMI matrices for the $ns\bar{c}\bar{c}$,
$ns\bar{b}\bar{b}$, and $ns\bar{c}\bar{b}$ systems.}\label{class56}
\begin{tabular}{cccc}\hline\hline
System&$J^P$&Wave Functions&$\langle H_{CM}\rangle$\\\hline

$ns\bar{c}\bar{c}$, $ns\bar{b}\bar{b}$&$0^+$&$(\phi_2\chi_3$,
$\phi_1\chi_6)^T$& $\left(\begin{array}{cc}
\frac{8}{3}(\alpha-\beta)  &2\sqrt{6}\beta\\
                           &4\alpha\\
\end{array}\right)$\\

&$1^+$&$(\phi_2\chi_5, \phi_1\chi_4, \phi_2\chi_2)^T$&
$\left(\begin{array}{ccc}
-\frac{8}{3}\theta  &-2\sqrt{2}\beta  &-\frac{4\sqrt{2}}{3}\gamma\\
                    &-\frac{4}{3}\eta &-4\gamma\\
                    &                 &\frac{4}{3}(2\alpha-\beta)
\end{array}\right)$\\

&$2^+$&$(\phi_2\chi_1)$&$\frac{4}{3}(2\alpha+\beta)$\\\\

$ns\bar{c}\bar{b}$&$0^+$&$(\phi_2\chi_6, \phi_1\chi_3, \phi_2\chi_3,
\phi_1\chi_6)^T$& $\left(\begin{array}{cccc}
-8\alpha    &2\sqrt{6}\beta                 &-\frac{4}{\sqrt{3}}\mu           &0\\
            &-\frac{4}{3}(\alpha+5\beta)   &4\sqrt{2}\mu                     &-\frac{10\mu}{\sqrt{3}}\\
            &                              &\frac{8}{3}(\alpha-\beta)        &2\sqrt{6}\beta\\
            &                              &                                 &4\alpha\\
\end{array}\right)$\\

&$1^+$&$(\phi_2\chi_5, \phi_1\chi_4, \phi_1\chi_2, \phi_2\chi_4,
\phi_1\chi_5$, $\phi_2\chi_2)^T$& $\left(\begin{array}{cccccc}
-\frac{8}{3}\theta &-2\sqrt{2}\beta  &4\delta                       &\frac{4}{3}\mu   &-\frac{4\sqrt{2}}{3}\gamma  &0\\
                   &-\frac{4}{3}\eta &\frac{10\sqrt{2}}{3}\delta    &0                &-4\gamma                    &\frac{10}{3}\mu\\
                   &                 &-\frac{2}{3}(2\alpha+5\beta)  &-4\gamma         &2\sqrt{2}\mu                &-\frac{10\sqrt{2}}{3}\gamma\\
                   &                 &                              &\frac{8}{3}\eta  &\frac{4\sqrt{2}}{3}         &-2\sqrt{2}\beta\\
                   &                 &                              &                 &\frac{4}{3}(2\alpha-\beta)  &4\delta\\
                   &                 &                              &                 &                            &\frac{4}{3}\theta\\
\end{array}\right)$\\

&$2^+$&$(\phi_1\chi_1, \phi_2\chi_1)^T$& $\left(\begin{array}{cc}
\frac{2}{3}(-2\alpha+5\beta)  &-2\sqrt{2}\mu\\
                              &\frac{4}{3}(2\alpha+\beta)\\
\end{array}\right)$
\\\hline\hline
\end{tabular}
\end{table*}

\section{NUMERICAL RESULTS}\label{sec3}

\subsection{Parameters}

We need to determine the values of the relevant coefficients
$C_{qq}$, $C_{\bar{Q}q}$, and $C_{QQ}$ in estimating the mass
splittings of the possible tetraquark states. Here, the subscripts
$Q=c$, $b$ and $q=n$, $s$ with $n=u$, $d$. The parameters
$C_{nn}=18.3$ MeV and $C_{ns}=12.3$ MeV can be determined from the
CMI relation between $N-\Delta$ and $\Sigma-\Sigma^*$ systems. One
further obtains $C_{ss}=6.4$ MeV by using the relation
$2M_{\Omega}+M_{\Delta}-(2M_{\Xi^*}+M_{\Xi})=8C_{ss}+8C_{nn}$. The
coupling constant $C_{\bar{c}n}$ ($C_{\bar{c}s}$) is estimated by
considering the mass splitting between the pseudoscalar and vector
charmed (charmed-strange) mesons. Similarly, $C_{\bar{b}n}$
($C_{\bar{b}s}$) is determined with the bottom (bottom-strange)
mesons. We extract the $C_{c\bar{c}}$ ($C_{b\bar{b}}$) from the
associated charmonium (bottomium) mesons and use the mass of $B_c^*$
estimated from GI model \cite{Godfrey:1985xj} to determine
$C_{c\bar{b}}$. For the remaining constants $C_{QQ}$ ($C_{cc}$,
$C_{bc}$, $C_{bb}$), however, there are no observed or confirmed
doubly heavy baryons in experiments at present. The approximation
$C_{QQ}=C_{Q\bar{Q}}$, i.e., $C_{c\bar{c}}=C_{cc}$,
$C_{c\bar{b}}=C_{cb}$, and $C_{b\bar{b}}=C_{bb}$, is used in our
analysis of the $qq\bar{Q}\bar{Q}$ tetraquarks. We collect the
determined parameters in Table \ref{parameters}. In getting these
parameters and in the following evaluation, we use the following
meson masses \cite{Olive:2016xmw}: $M_D=1867.21$ MeV,
$M_{D^*}=2008.56$ MeV, $M_{D_s}=1968.27$ MeV, $M_{D_s^{*}}=2112.1$
MeV, $M_{B}=5279.31$ MeV, $M_{B^{*}}=5324.65$ MeV, $M_{B_s}=5366.82$
MeV, and $M_{B_s^{*}}=5415.4$ MeV.

\begin{table*}[htbp]\centering
\caption{The parameters obtained with mass splittings of
conventional mesons and baryons}\label{parameters} \centering
\begin{tabular}{ccccc}\hline\hline
Hadron&CMI&Hadron&CMI&Parameter (MeV)\\\hline
$N$      &$-8C_{nn}$                             &$\Delta$                      &$8C_{nn}$                              &$C_{nn}=18.3$      \\
$\Sigma$ &$\frac{8}{3}C_{nn}-\frac{32}{3}C_{ns}$ &$\Sigma^{*}$                  &$\frac{8}{3}C_{nn}+\frac{16}{3}C_{ns}$ &$C_{ns}=12.3$      \\
$D$      &$-16C_{\bar{c}n}$                      &$D^*$                         &$\frac{16}{3}C_{\bar{c}n}$             &$C_{\bar{c}n}=6.6$ \\
$D_s$    &$-16C_{\bar{c}s}$                      &$D^*_s$                       &$\frac{16}{3}C_{\bar{c}s}$             &$C_{\bar{c}s}=6.7$ \\
$B$      &$-16C_{\bar{b}n}$                      &$B^*$                         &$\frac{16}{3}C_{\bar{b}n}$             &$C_{\bar{b}n}=2.1$ \\
$B_s$    &$-16C_{\bar{b}s}$                      &$B^*_s$                       &$\frac{16}{3}C_{\bar{b}s}$             &$C_{\bar{b}s}=2.3$ \\
$\eta_c$ &$-16C_{\bar{c}c}$                      &$J/\psi$                      &$\frac{16}{3}C_{\bar{c}c}$             &$C_{\bar{c}c}=5.3$ \\
$\eta_b$ &$-16C_{\bar{b}b}$                      &$\Upsilon$                    &$\frac{16}{3}C_{\bar{b}b}$             &$C_{\bar{b}b}=2.9$ \\
$B_c$    &$-16C_{\bar{c}b}$                      &$B^*_c$ \cite{Godfrey:1985xj}  &$\frac{16}{3}C_{\bar{c}b}$             &$C_{\bar{c}b}=3.3$ \\
\hline\hline
\end{tabular}
\end{table*}

Since we only consider the color magnetic interaction between
valence quarks in a $qq\bar{Q}\bar{Q}$ system and no dynamical
effects are involved in this calculation, we will discuss the masses
of tetraquarks with two schemes. In the first scheme, we use the
experimental data to obtain the effective constituent quark masses
and estimate the values of the tetraquark masses by introducing the
mass shifts due to CMI. The corresponding formula is
$M=\sum\limits_{i}m_i+\langle H_{CM}\rangle$. In Table
\ref{parametermass}, we show the procedure of determination. In the
second scheme, we relate the tetraquark system to a reference
meson-meson system and compare the masses to the threshold of the
meson-meson state by the equation $M=M_{\rm ref}-\langle
H_{CM}\rangle_{\rm ref}+\langle H_{CM}\rangle$. In the first scheme,
we find that the predicted masses are generally overestimated. The
main reason is probably from the fact that the dynamical effects can
not be simply absorbed into the effective quark masses. This scheme
can give an upper limit for the masses of the tetraquark states. In
the following discussions, we mainly focus on the results obtained
from the threshold scheme. By substituting the $C_{ij}$ parameters
into the corresponding CMI matrices given in the previous section,
one gets the values of the mass shifts from the color magnetic
interaction after we diagonalize the numerical CMI matrices.

\begin{table}[htbp]
\caption{The effective constituent quark masses extracted from
conventional baryons}\label{parametermass}
\begin{tabular}{ll}\hline\hline
Mass formula& Quark mass (MeV)\\\hline
$M_N=3m_n-8C_{nn}$&$m_n=361.8$\\
$M_{\Omega}=3m_s+8C_{ss}$&$m_s=540.4$\\
$M_{\Sigma_c}=\frac{8}{3}C_{nn}-\frac{32}{3}C_{nc}+2m_{q}+m_{c}$&\multirow{2}{*}{$m_c=1724.8$}\\
$M_{\Sigma_c^*}=\frac{8}{3}C_{nn}+\frac{16}{3}C_{nc}+2m_{n}+m_{c}$\\
$M_{\Sigma_b}=2m_{q}+m_{b}+\frac{8}{3}C_{nn}-\frac{32}{3}C_{bn}$&\multirow{2}{*}{$m_{b}=5052.9$}\\
$M_{\Sigma_b^*}=2m_{q}+m_{b}+\frac{8}{3}C_{nn}+\frac{16}{3}C_{bn}$\\
\hline\hline
\end{tabular}
\end{table}

 Since the corresponding antiparticle $\bar{q}\bar{q}QQ$ has identical mass to the $qq\bar{Q}\bar{Q}$, we only present the results for the $qq\bar{Q}\bar{Q}$ systems. In the previous section, we have divided such systems into six classes. In the following part, we discuss the systems according to their $SU(3)$ classification in the flavor space.

\subsection{Systems with strangeness=0}

\begin{table*}[htbp]\centering
\caption{Numerical results for the $nn\bar{c}\bar{c}$,
$nn\bar{b}\bar{b}$, and $nn\bar{c}\bar{b}$ systems in units of MeV,
where $n=u$ or $d$. The masses in the sixth column are estimated
with the masses of effective quarks and those in the last column
with the thresholds of $DD$/$BB$/$BD$.}\label{tablennQQ}
\begin{tabular}{ccccccc}\hline\hline
System&$J^P$&$\langle
H_{CM}\rangle$&Eigenvalues&Eigenvectors&Mass&$DD$/$BB$/$BD$\\\hline
$(nn\bar{c}\bar{c})^{I=1}$ &$0^+$&
$\left(\begin{array}{cc}-7.5&129.3\\129.3&94.4\end{array}\right)$&
$\left[\begin{array}{c}182.5\\-95.5\end{array}\right]$&
$\left[\begin{array}{c}(0.56,0.83)\\(-0.83,0.56)\end{array}\right]$&
$\left[\begin{array}{c}4356\\4078\end{array}\right]$&
$\left[\begin{array}{c}4128\\3850\end{array}\right]$\\

&$1^+$&27.7&27.7&1.00&4201&3973\\
&$2^+$&98.1&98.1&1.00&4271&4044\\

$(nn\bar{c}\bar{c})^{I=0}$ &$1^+$&
$\left(\begin{array}{cc}-132.3&-74.7\\-74.7&-3.2\end{array}\right)$&
$\left[\begin{array}{c}-166.4\\31.0\end{array}\right]$&
$\left[\begin{array}{c}(0.91,0.42)\\(-0.42,0.91)\end{array}\right]$&
$\left[\begin{array}{c}4007\\4204\end{array}\right]$&
$\left[\begin{array}{c}3779\\3977\end{array}\right]$\\\hline
$(nn\bar{b}\bar{b})^{I=1}$&$0^+$&
$\left(\begin{array}{cc}34.1&41.2\\41.2&84.8\end{array}\right)$&
$\left[\begin{array}{c}107.8\\11.1\end{array}\right]$&
$\left[\begin{array}{c}(0.49,0.87)\\(-0.87,0.49)\end{array}\right]$&
$\left[\begin{array}{c}10937\\10841\end{array}\right]$&
$\left[\begin{array}{c}10734\\10637\end{array}\right]$\\
&$1^+$&45.3&45.3&1.00&10875&10671\\
&$2^+$&67.7&67.7&1.00&10897&10694\\

$(nn\bar{b}\bar{b})^{I=0}$&$1^+$&
$\left(\begin{array}{cc}-138.7&-23.8\\-23.8&-12.8\end{array}\right)$&
$\left[\begin{array}{c}-143.0\\-8.5\end{array}\right]$&
$\left[\begin{array}{c}(0.98,0.18)\\(-0.18,0.98)\end{array}\right]$&
$\left[\begin{array}{c}10686\\10821\end{array}\right]$&
$\left[\begin{array}{c}10483\\10617\end{array}\right]$\\
\hline $(nn\bar{c}\bar{b})^{I=1}$ &$0^+$&
$\left(\begin{array}{cc}11.2&85.2\\85.2&86.4\end{array}\right)$&
$\left[\begin{array}{c}142.0\\-44.4\end{array}\right]$&
$\left[\begin{array}{c}(0.55,0.84)\\(0.84,-0.55)\end{array}\right]$&
$\left[\begin{array}{c}7643\\7457\end{array}\right]$&
$\left[\begin{array}{c}7428\\7241\end{array}\right]$\\

&$1^+$&
$\left(\begin{array}{ccc}22.4&17.0&-49.2\\17.0&34.4&36.0\\-49.2&36.0&68.8\end{array}\right)$&
$\left[\begin{array}{c}46.4\\-28.0\\107.2\end{array}\right]$&
$\left[\begin{array}{c}(0.59,0.81,-0.01)\\(0.69,-0.49,0.53)\\(-0.43,0.32,0.85)\end{array}\right]$&
$\left[\begin{array}{c}7548\\7473\\7609\end{array}\right]$&
$\left[\begin{array}{c}7332\\7258\\7393\end{array}\right]$\\

&$2^+$&80.8&80.8&1.00&7582&7367\\

$(nn\bar{c}\bar{b})^{I=0}$ &$0^+$&
$\left(\begin{array}{cc}-172.8&85.2\\85.2&-144.8\end{array}\right)$&
$\left[\begin{array}{c}-245.2\\-72.4\end{array}\right]$&
$\left[\begin{array}{c}(-0.76,0.65)\\(0.65,0.76)\end{array}\right]$&
$\left[\begin{array}{c}7256\\7429\end{array}\right]$&
$\left[\begin{array}{c}7041\\7213\end{array}\right]$\\

&$1^+$&
$\left(\begin{array}{ccc}-137.6&-49.2&36.0\\-49.2&-11.2&42.4\\36.0&42.4&-86.8\end{array}\right)$&
$\left[\begin{array}{c}-180.2\\14.9\\-70.3\end{array}\right]$&
$\left[\begin{array}{c}(-0.81,-0.35,0.47)\\(0.23,-0.92,-0.31)\\(0.54,-0.14,0.83)\end{array}\right]$&
$\left[\begin{array}{c}7321\\7516\\7431\end{array}\right]$&
$\left[\begin{array}{c}7106\\7301\\7215\end{array}\right]$\\

&$2^+$&29.2&29.2&1.00&7530&7315\\\hline\hline
\end{tabular}
\end{table*}

\begin{figure*}[!h]
\centering
\begin{tabular}{ccccc}
\includegraphics[width=0.3\textwidth]{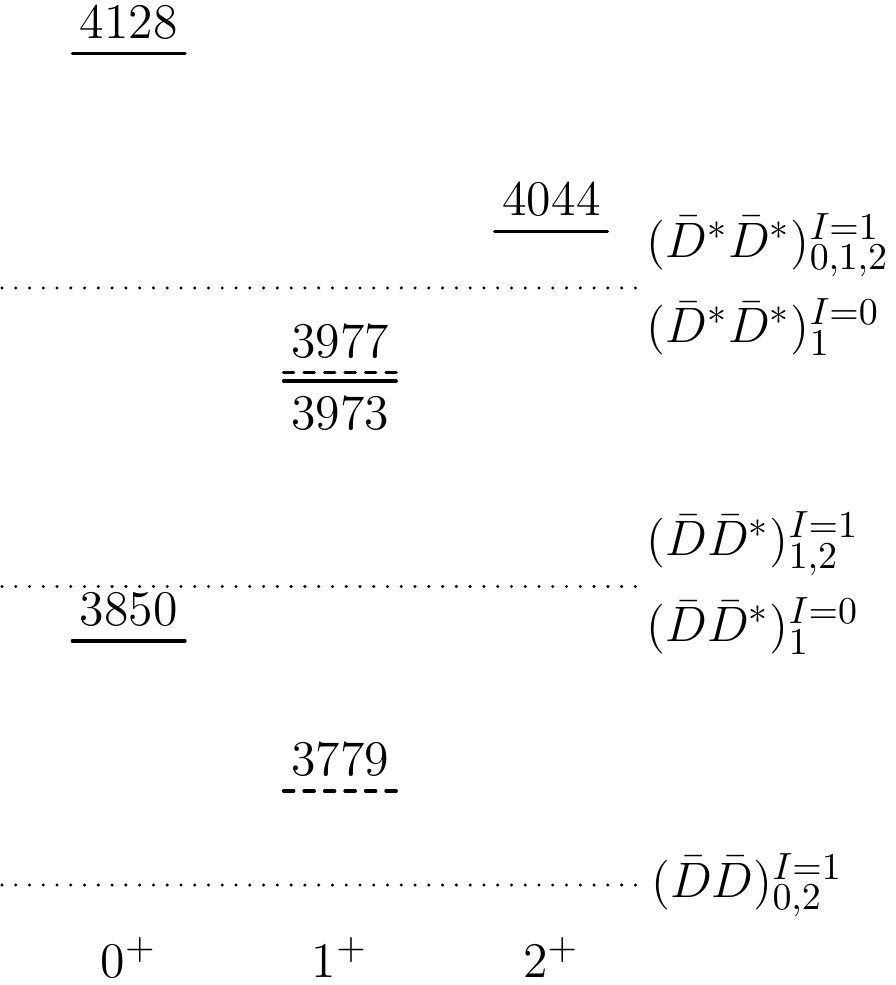}&\qquad
\includegraphics[width=0.3\textwidth]{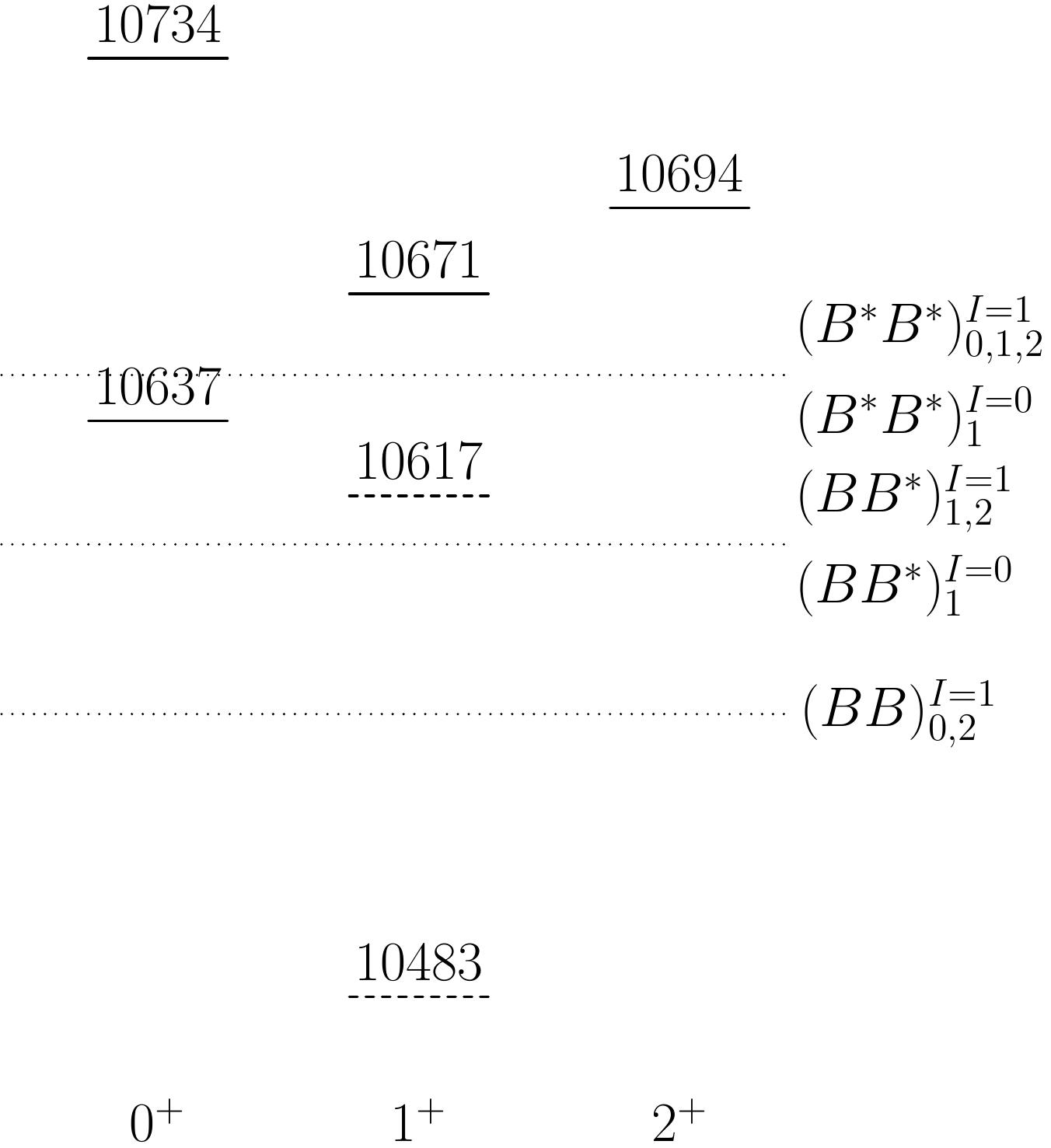}&\qquad
\includegraphics[width=0.3\textwidth]{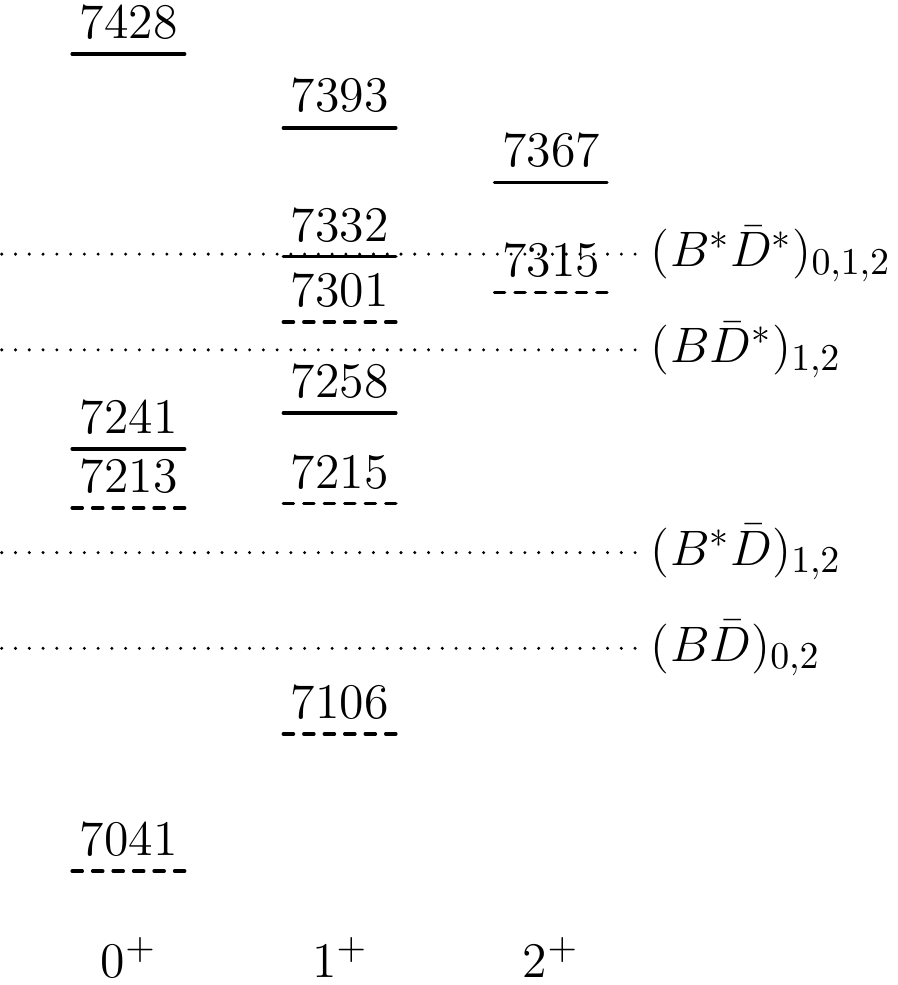}
\end{tabular}
\caption{The estimated masses (units: MeV) of the $nn\bar{c}\bar{c}$
(left), $nn\bar{b}\bar{b}$ (middle), and $nn\bar{c}\bar{b}$ (right)
systems in the threshold scheme. The solid (dashed) lines correspond
to the $I=1$ ($I=0$) case. The dotted lines are thresholds of the
possible decay channels. When the isospin conservation is satisfied
and the total spin of a tetraquark is equal to a subscript of the
symbol for a meson-meson channel, the corresponding $S$- or $D$-wave
decay is allowed.}\label{fignnQQ}
\end{figure*}

The quark content of these systems may be $nn\bar{c}\bar{c}$,
$nn\bar{b}\bar{b}$, or $nn\bar{c}\bar{b}$. We present the calculated
CMI matrices, the eigenvectors, the eigenvalues, and the estimated
masses in both schemes in Table \ref{tablennQQ}. As mentioned
earlier, the constitute quark scheme gives higher mass predictions.
In the following discussions, we will use the threshold scheme to
discuss their properties. In Figure \ref{fignnQQ}, the rough
positions of the studied tetraquark states, various thresholds, and
relevant rearrangement decay patterns are shown.

\subsubsection{$nn\bar{c}\bar{c}$}

In literature, a compact tetraquark state with the quark content
$nn\bar{c}\bar{c}$ is usually called $T_{cc}$. Because of the highly
symmetric constraint from the Pauli principle, the total number of
the $T_{cc}$ states is restricted to be six: four isovector states
and two isoscalar states. In the isovector case, only one scalar
state has probably attractive color-magnetic interaction. The mixing
with the other scalar state makes the gap between the two states
larger, from 100 MeV to 280 MeV (see Tab. \ref{tablennQQ}), which
results in the heaviest $T_{cc}$ and the lightest $0^+$ $T_{cc}$. In
the isoscalar case, the two CMI matrix elements for the axial-vector
states can be both negative, depending on the parameters. The state
mixing finally leads to one tetraquark with attractive CMI and one
with repulsive CMI. This lower tetraquark corresponds to the lowest
$T_{cc}$ which was widely discussed in literature. In Ref.
\cite{Hyodo:2012pm}, we argued that the tetraquark mixing is
suppressed because the transition needs a spin-flip for the $cc$
state. Now, one sees that the mass shift for the $I=0$ tetraquarks
due to the state mixing is about 30 MeV and it is not a big number.
On the other hand, the mass shift for the isovector $0^+$ states due
to the state mixing is about 90 MeV and the state mixing may have
effects on the decay widths.

From Fig. \ref{fignnQQ}, the lowest $T_{cc}$ should be stable if our
estimation is reasonable. It is below the $\bar{D}\bar{D}^*$
threshold and its strong decay into $\bar{D}\bar{D}$ is forbidden by
both isospin and angular momentum conservations. Whether it can be
observed would be crucial information in distinguishing models of
genuine multiquark states, although there is still no experimental
signal for this state. If the other isoscalar $1^+$ state is narrow
\cite{Hyodo:2012pm}, the isovector $1^+$ $T_{cc}$ might also be
narrow since they have similar masses and rearrangement decay
patterns. Another interesting state is the tensor $T_{cc}$ with
$I=1$. Its decay into $\bar{D}\bar{D}$ or $\bar{D}\bar{D}^*$ is
through $D$ wave and relevant partial widths should not be large.
Although this tetraquark can decay into $\bar{D}^*\bar{D}^*$ through
$S$ wave, the width may not be so large because of the suppression
in phase space. This feature of narrow width for high-spin state is
similar to the $J=\frac52$ pentaquark case which was studied in Ref.
\cite{Wu:2017weo}. The lower $0^+$ $T_{cc}$ has only one $S$-wave
decay channel $\bar{D}\bar{D}$ while the higher one can also decay
into $\bar{D}^*\bar{D}^*$ through $S$ or $D$ wave. The decay into
the channel $\bar{D}\bar{D}^*$ is forbidden either by kinematics or
by angular momentum conservation. Probably the higher scalar
$T_{cc}$ could not be observed even if it really exists because of
its broad width. From our estimations, probably a search in the
$I=1$ $\bar{D}^{(*)}\bar{D}^{(*)}$ channel may show us some exotic
signals.

\subsubsection{$nn\bar{b}\bar{b}$}

Similarly, such a system is called $T_{bb}$. From Tab.
\ref{class12}, the basic features of spectrum should be similar to
the $T_{cc}$ case. Because the concrete interaction strengthes are
different, however, the relevant positions of $T_{bb}$'s are changed
compared with their charmed partners. Accordingly, the decay
properties are different. We now take a look at the tetraquark
properties from Fig. \ref{fignnQQ}. Below the $BB^*$ threshold,
there is only one state, the stable $1^+$ $T_{bb}$ with $I=0$. The
second heavier state is the other $1^+$ $T_{bb}$ with $I=0$ which is
just above the $BB^*$ threshold and should be narrower than its
charmed analog. The remaining $T_{bb}$ states are all isovector
mesons and are all above these two states. Although the lowest $0^+$
$T_{bb}$ is above the $BB^*$ threshold, its decay into this channel
is forbidden and the width may be comparable to the lowest $0^+$
$T_{cc}$. Now, the $I(J^P)=1(1^+)$ $T_{bb}$ is above the $B^*B^*$
threshold. Because the decay into this channel through $S$ wave is
forbidden, its dominant decays should be $S$-wave $BB^*$, $D$-wave
$BB^*$, and $D$-wave $B^*B^*$ and thus the width is larger than the
$I(J^P)=0(1^+)$ $T_{bb}$. The features of the tensor $T_{bb}$ and
the heaviest $T_{bb}$ are similar to their charmed analogs,
respectively.

\subsubsection{$nn\bar{c}\bar{b}$}

Now we focus on the systems with two identical light quarks but
different heavy quarks. We use $T_{cb}$ to denote such tetraquarks.
Obviously, the wave functions of the heavy antidiquarks are not
constrained by the Pauli principle and the number of states is
bigger than that in the $nn\bar{c}\bar{c}$ systems. From Tab.
\ref{class34} and Tab. \ref{class12}, by comparing $T_{cb}$ with the
$T_{cc}$ in the $I=1$ case, we have the same number of $0^+$ and
$2^+$ tetraquark states and two more $1^+$ states. In the isoscalar
case, two $0^+$ states, one $2^+$ state, and one more $1^+$ state
are allowed.

The heaviest and the lightest states in the $I=1$ case are the two
$0^+$ $T_{cb}$'s, whose mass difference is about 180 MeV. The masses
of the $1^+$ and $2^+$ states lie between these two tetraquarks. The
$I=0$ $T_{cb}$ masses are generally lower than those $I=1$ masses.
Different from the $T_{cc}$ and $T_{bb}$ mesons, now the lowest
state is a scalar tetraquark.

From Fig. \ref{fignnQQ}, there are four possible rearrangement decay
channels for the $T_{cb}$ states, $B\bar{D}$, $B^*\bar{D}$,
$B\bar{D}^*$, and $B^*\bar{D}^*$, two of which are
pseudoscalr+vector (PV) type channels. Since no symmetry constraint
is required among isospin, spin, and orbital spaces for these
meson-meson states, each $T_{cb}$ can decay into these channels once
the angular momentum conservation is satisfied and the mass is high
enough. We do not need to consider the isospin conservation in
considering the meson-meson decay properties.

Two states, one scalar meson and one axial vector meson, are below
the threshold of $B\bar{D}$ and they should be narrow and stable
tetraquarks. Two higher $0^+$ states can only decay into $B\bar{D}$
through $S$ wave although they are above the threshold of
$B^*\bar{D}$. The highest $0^+$ state may be broad and might not be
observed since it can decay into $B^*\bar{D}^*$ through $S$ and $D$
wave and into $B\bar{D}$ through $S$ wave. Each excited $1^+$
$T_{cb}$ has the $S$-wave decay channel $B^*\bar{D}$. The $S$-wave
channel $B\bar{D}^*$ is opened for three of them and the $S$-wave
$B^*\bar{D}^*$ is also opened for the highest one. For the two
tensor states, the lower one can only decay through $D$ wave and the
higher one can also decay into $B^*\bar{D}^*$ through $S$ wave.
Needless to say, these arguments strongly depend on the estimation
of the $T_{cb}$ masses, which is the dominant uncertainty of the
present method. If our estimation is reasonable, probably exotic
states could be observed in the $B^{(*)}\bar{D}^{(*)}$ channels. If
our estimation is underestimated around 100 MeV, the lowest two
states may be above respective $S$-wave decay channels and their
signals may also be observed through two-body strong decays. Anyway,
interesting phenomena may exist in the $B^{(*)}\bar{D}^{(*)}$
channels.

\newsavebox{\tablebox}
\begin{table*}[htbp]\centering\tiny
\caption{Numerical results for the $ns\bar{c}\bar{c}$, $ns\bar{b}b$,
and $ns\bar{c}\bar{b}$ systems in units of MeV, where $n=u$ or $d$.
The masses in the sixth column are estimated with the masses of
effective quarks and those in the last column with the thresholds of
$DD_s$/$BB_s$/$DB_s$. For the $ns\bar{c}\bar{b}$ system, the result
estimated with the threshold of $BD_s$ is about 10 MeV higher than
that with $DB_s$.}\label{tablensQQ}
\begin{tabular}{ccccccc}\hline\hline
System&$J^P$&$\langle
H_{CM}\rangle$&Eigenvalues&Eigenvectors&Mass&$DD_s$/$BB_s$/$DB_s$\\\hline

$ns\bar{c}\bar{c}$&$0^+$&
$\left(\begin{array}{cc}-24.0&130.3\\130.3&70.4\end{array}\right)$&
$\left[\begin{array}{c}161.8\\-115.4\end{array}\right]$&
$\left[\begin{array}{c}(0.57,0.82)\\(-0.82,0.57)\end{array}\right]$&
$\left[\begin{array}{c}4514\\4236\end{array}\right]$&
$\left[\begin{array}{c}4210\\3933\end{array}\right]$\\

&$1^+$&
$\left(\begin{array}{ccc}-84.3&-75.2&0.4\\-75.2&4.8&0.8\\0.4&0.8&11.5\end{array}\right)$&
$\left[\begin{array}{c}-127.2\\47.7\\11.5\end{array}\right]$&
$\left[\begin{array}{c}(-0.87,-0.50,0.01)\\(0.50,-0.87,-0.01)\\(0.01,-0.01,1.0)\end{array}\right]$&
$\left[\begin{array}{c}4225\\4400\\4363\end{array}\right]$&
$\left[\begin{array}{c}3921\\4096\\4060\end{array}\right]$\\

&$2^+$&82.4&82.4&1.00&4434&4131\\\hline
$ns\bar{b}\bar{b}$ &$0^+$&
$\left(\begin{array}{cc}17.1&43.1\\43.1&60.8\end{array}\right)$&
$\left[\begin{array}{c}87.3\\-9.4\end{array}\right]$&
$\left[\begin{array}{c}(0.52,0.85)\\(-0.85,0.52)\end{array}\right]$&
$\left[\begin{array}{c}11095\\10999\end{array}\right]$&
$\left[\begin{array}{c}10804\\10707\end{array}\right]$\\

&$1^+$&
$\left(\begin{array}{ccc}-90.7&-24.9&0.8\\-24.9&-4.8&1.6\\0.8&1.6&28.8\end{array}\right)$&
$\left[\begin{array}{c}-97.4\\28.9\\1.8\end{array}\right]$&
$\left[\begin{array}{c}(0.97,0.26,-0.01)\\(0.00,0.05,1.0)\\(0.26,-0.96,0.05)\end{array}\right]$&
$\left[\begin{array}{c}10911\\11037\\11010\end{array}\right]$&
$\left[\begin{array}{c}10619\\10745\\10718\end{array}\right]$\\

&$2^+$&52.3&52.3&1.00&11060&10769\\
\hline $ns\bar{c}\bar{b}$ &$0^+$&
$\left(\begin{array}{cccc}-124.8&86.7&-0.2&0.0\\86.7&-138.8&0.6&-0.6\\-0.2&0.6&-5.6&86.7\\0.0&-0.6&86.7&62.4\end{array}\right)$&
$\left[\begin{array}{c}-218.8\\121.5\\-64.7\\-44.8\end{array}\right]$&
$\left[\begin{array}{c}(0.68,-0.74,0.00,0.00)\\(0.00,0.00,0.56,0.83)\\(0.02,0.01,-0.83,0.56)\\(0.73,0.68,0.02,-0.01)\end{array}\right]$&
$\left[\begin{array}{c}7461\\7801\\7615\\7635\end{array}\right]$&
$\left[\begin{array}{c}7158\\7498\\7312\\7332\end{array}\right]$\\

&$1^+$&
$\left(\begin{array}{cccccc}-89.6&-50.1&35.6&0.1&0.6&0.0\\-50.1&-3.2&42.0&0.0&1.2&0.3\\35.6&42.0&-79.8&1.2&0.3&1.4\\0.1&0.0&1.2&6.4&16.8&-50.1\\0.6&1.2&0.3&16.8&18.0&35.6\\0.0&0.3&1.4&-50.1&35.6&44.8\end{array}\right)$&
$\left[\begin{array}{c}-149.8\\86.0\\-49.2\\-46.4\\30.2\\25.7\end{array}\right]$&
$\left[\begin{array}{c}(0.69,0.41,-0.60,0.01,-0.01,0.01)\\(0.00,0.01,0.01,-0.45,0.32,0.83)\\(-0.60,-0.01,-0.70,0.26,-0.17,0.21)\\(0.24,0.00,0.29,0.62,-0.45,0.51)\\(0.06,-0.20,-0.07,-0.57,-0.79,0.00)\\(0.31,-0.89,-0.25,0.12,0.18,0.01)\end{array}\right]$&
$\left[\begin{array}{c}7530\\7766\\7631\\7634\\7710\\7706\end{array}\right]$&
$\left[\begin{array}{c}7227\\7462\\7327\\7330\\7407\\7402\end{array}\right]$\\

&$2^+$&
$\left(\begin{array}{cc}38.2&-0.3\\-0.3&65.2\end{array}\right)$&
$\left[\begin{array}{c}65.2\\38.2\end{array}\right]$&
$\left[\begin{array}{c}(-0.01,1.0)\\(-1.0,-0.01)\end{array}\right]$&
$\left[\begin{array}{c}7745\\7718\end{array}\right]$&
$\left[\begin{array}{c}7442\\7415\end{array}\right]$\\\hline\hline
\end{tabular}
\end{table*}

\begin{figure*}[htbp]\centering
\begin{tabular}{ccccc}
\includegraphics[width=0.3\textwidth]{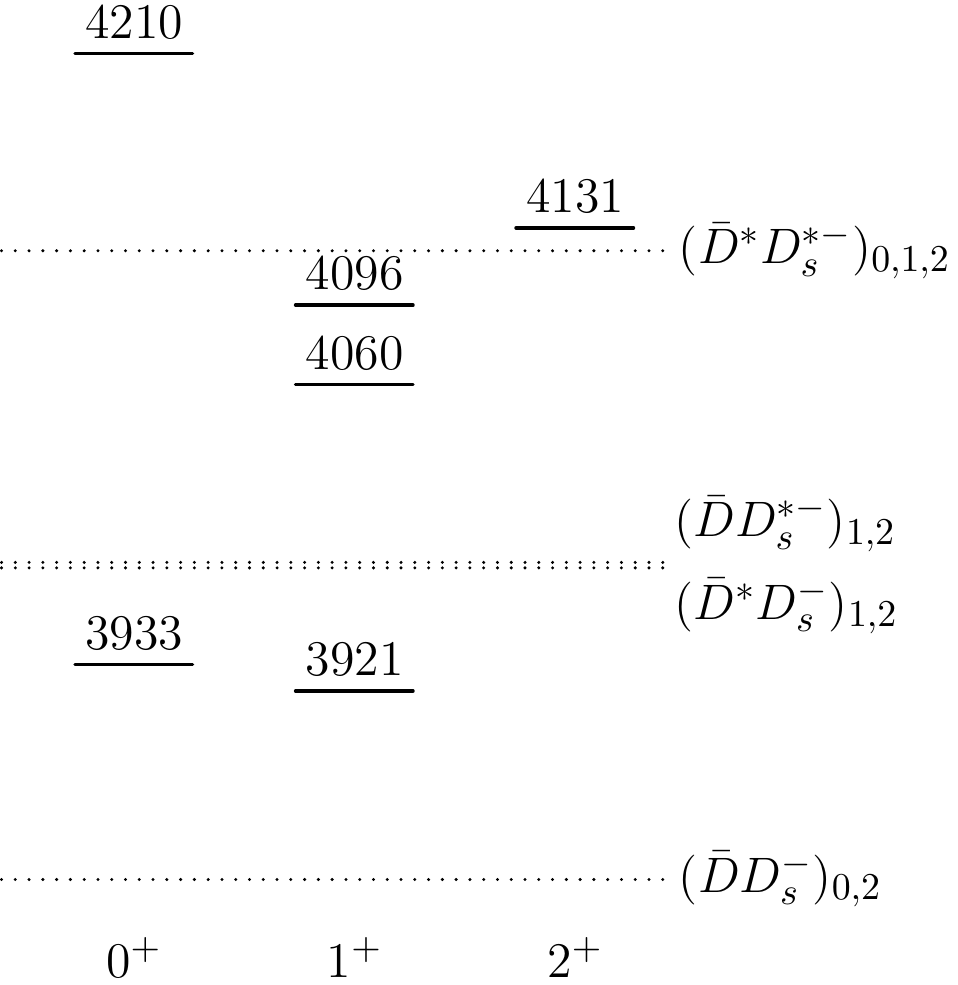}&\qquad
\includegraphics[width=0.3\textwidth]{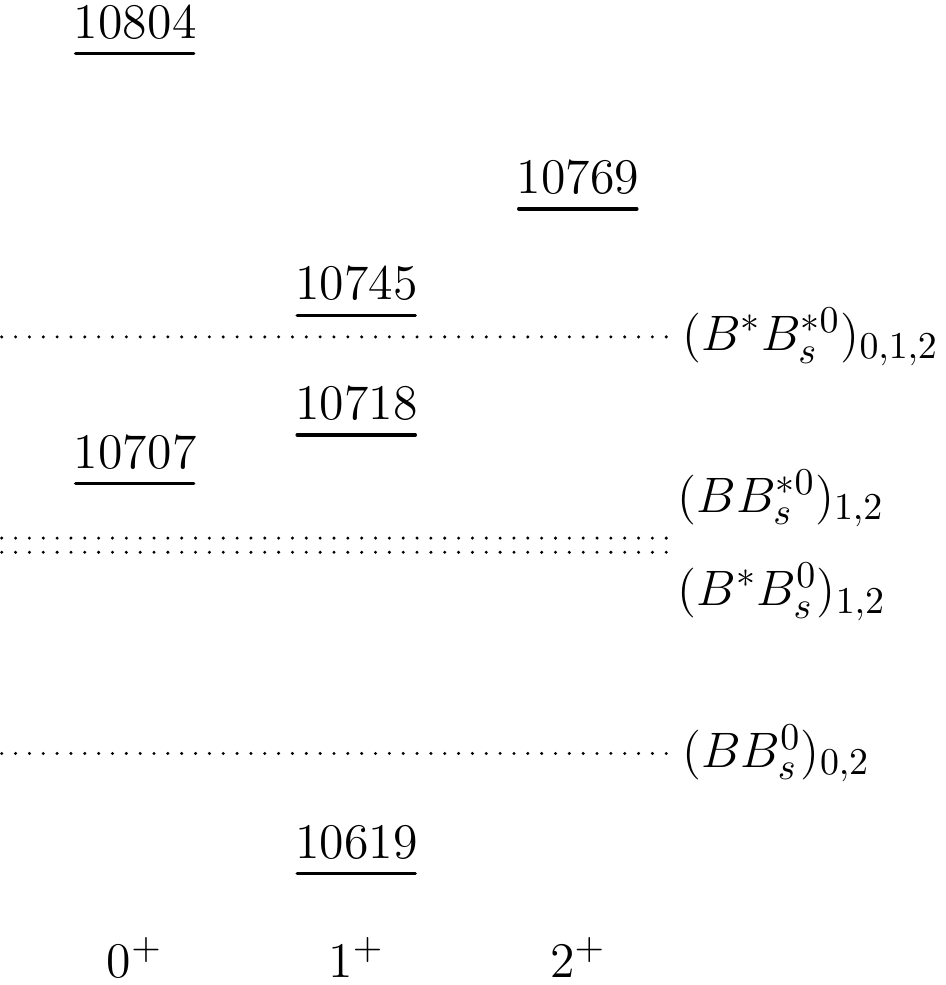}&\qquad
\includegraphics[width=0.3\textwidth]{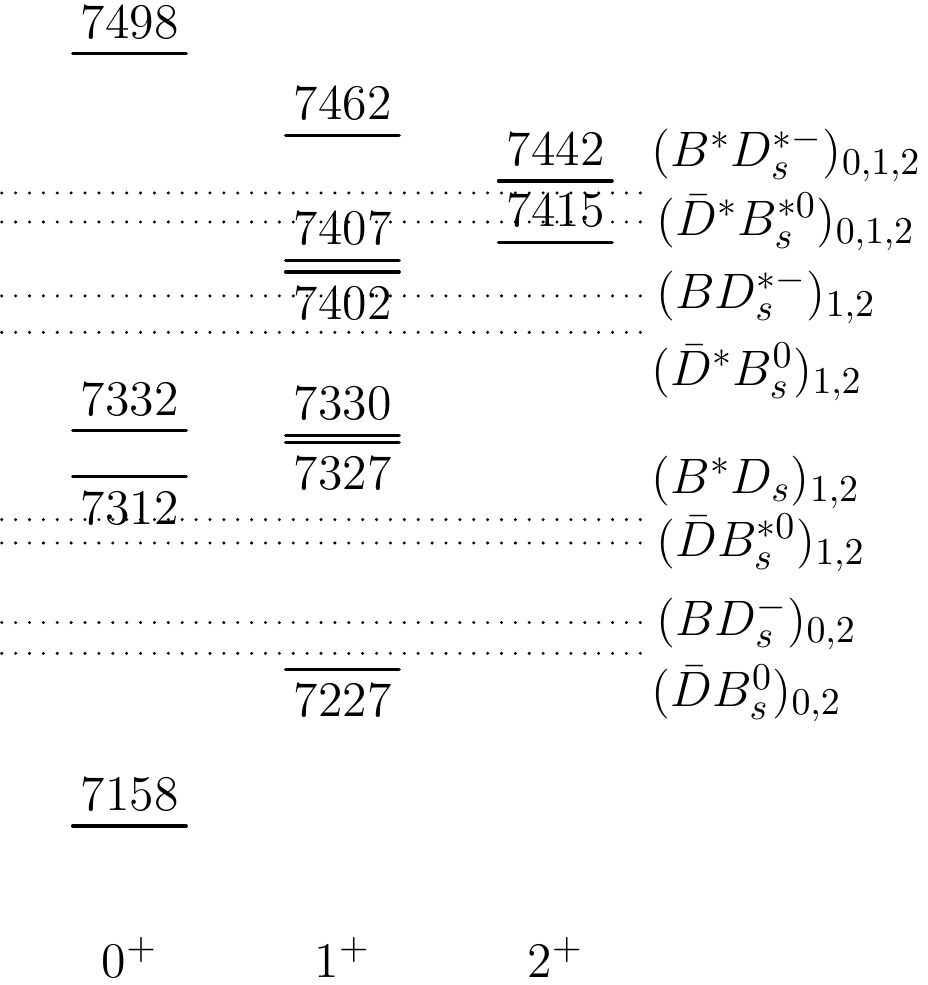}
\end{tabular}
\caption{The estimated masses (units: MeV) of the $ns\bar{c}\bar{c}$
(left), $ns\bar{b}\bar{b}$ (middle), and $ns\bar{c}\bar{b}$ (right)
systems in the threshold scheme. The dotted lines are thresholds of
the possible decay channels. When the total spin of a tetraquark is
equal to a subscript of the symbol for a meson-meson channel, the
corresponding $S$- or $D$-wave decay is allowed.}\label{fignsQQ}
\end{figure*}

\subsection{Systems with strangeness=$-1$}

The system $ns\bar{c}\bar{c}$, $ns\bar{b}\bar{b}$, or
$ns\bar{c}\bar{b}$ contains a strange quark and the isospin of each
state is $I=\frac12$. We list the numerical results in Tab.
\ref{tablensQQ} and show their rough positions in Fig.
\ref{fignsQQ}. In the table, one finds that some nondiagonal elements of $\langle H_{CM}\rangle$ are close to zero. They are from $\gamma=(C_{13}-C_{23})+(C_{14}-C_{24})$ and $\mu=(C_{13}-C_{23})-(C_{14}-C_{24})$ and their nonvanishing values reflect the effects of the flavor symmetry breaking. Comparing with the previous $nn\bar{Q}\bar{Q}$
systems, the total number of states does not change. Although the
constraint from isospin conservation is removed, we find that
$ns\bar{Q}\bar{Q}$ and $nn\bar{Q}\bar{Q}$ have some similar
properties of rearrangement decays.

First, we concentrate on the $ns\bar{c}\bar{c}$ systems. From Fig.
\ref{fignsQQ}, it is clear that the heaviest state is still a $0^+$
tetraquark and the lightest one is still a $1^+$ meson. This lowest
$1^+$ state does not decay into $\bar{D}D_s^-$ and this highest
$0^+$ state can decay into $\bar{D}D_s^-$ and $\bar{D}^*D_s^{*-}$
through $S$ wave, which is similar to corresponding states in Fig.
\ref{fignnQQ}. Similar features of decay properties for other mesons
also exist: the lower $0^+$ state decays into $\bar{D}D_s^-$ through
$S$ wave, the two higher $1^+$ states have PV type $S$-wave decay
channels, and the $2^+$ state might be a narrow tetraquark.

Next, we move on to the $ns\bar{b}\bar{b}$ systems. The positions of
masses for the $ns\bar{b}\bar{b}$ states and those of various
thresholds are similar to those in the $nn\bar{b}\bar{b}$ case. The
only difference is that the positions for the lower $0^+$ state and
the second $1^+$ state are exchanged. The similarities in mass
spectrum would result in those in widths.

Finally, we take a look at the systems composed of four different
flavors, $ns\bar{c}\bar{b}$. The lowest state is a stable scalar
tetraquark and the lowest $1^+$ is also below the corresponding
open-charm decay channel, which is a similar feature to the
$nn\bar{c}\bar{b}$ case. The decay properties of other states in the
$ns\bar{c}\bar{b}$ case and in the $nn\bar{c}\bar{b}$ case should
have similar features, too. A slightly different feature is that
almost degenerate states exist in the $1^+$ $ns\bar{c}\bar{b}$ case.
There are two states around 7330 MeV and two states around 7405 MeV.
By inspecting the contributions in the wave functions, one finds
that the dominant wave functions in one state belong to $I=1$
$nn\bar{c}\bar{b}$ and those in the other state belong to $I=0$
$nn\bar{c}\bar{b}$. It is the state mixing that results in the
occasional degeneracy.

When we estimate the $ns\bar{c}\bar{b}$ masses by using the
reference threshold $BD_s$ or $DB_s$, the mass difference is around
10 MeV. This value is from the $SU(3)$ flavor symmetry breaking.
From $M=M_{ref}-\langle H_{CM}\rangle_{ref}+\langle H_{CM}\rangle$
and Tab. \ref{parameters}, the difference resulting from the two
thresholds is
$[(M_{D_s}-M_D)-(M_{B_s}-M_B)]+16[(C_{\bar{c}n}-C_{\bar{c}s})-(C_{\bar{b}n}-C_{\bar{b}s})]$.
If the $SU(3)$ flavor symmetry is strict, $M_{D_s}=M_D$,
$M_{B_s}=M_B$, $C_{\bar{c}n}=C_{\bar{c}s}$,
$C_{\bar{b}n}=C_{\bar{b}s}$, and the difference vanishes. One may
also estimate the $ns\bar{Q}\bar{Q}$ masses with the PV type
thresholds and similar uncertainty occurs because of the $SU(3)$
symmetry breaking. For example, the difference resulting from the
thresholds $BB_s^{*0}$ and $B^*B_s^0$ is
$[(M_{B_s^*}-M_{B_s})-(M_{B_s}-M_B)+\frac{64}{3}(C_{\bar{b}n}-C_{\bar{b}s})]$,
which vanishes in the limit $m_n=m_s$. In fact, what the above
mentioned similarity reflects is also the underlying $SU(3)$ flavor
symmetry. More results will be discussed later.

\subsection{Systems with strangeness=$-2$}

\begin{figure*}[htbp]\centering
\begin{tabular}{ccccc}
\includegraphics[width=0.3\textwidth]{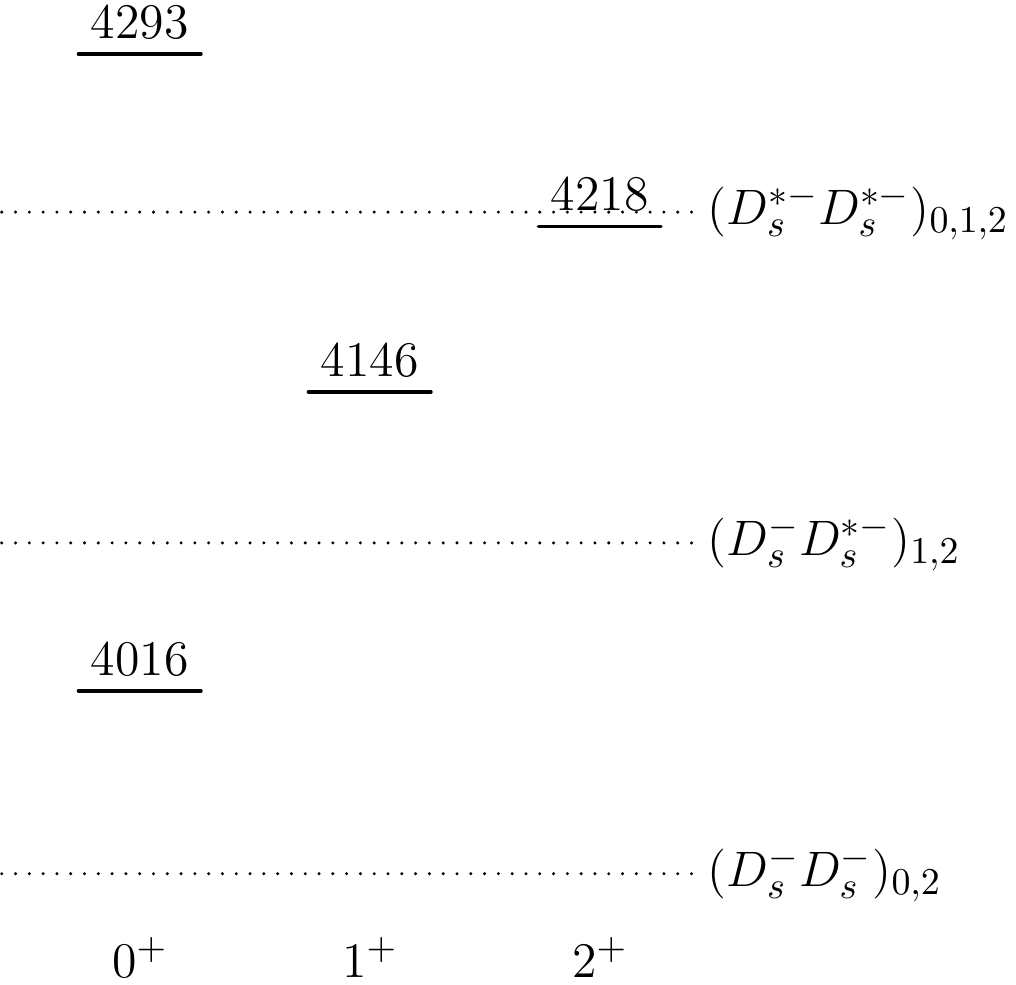}&\qquad&
\includegraphics[width=0.3\textwidth]{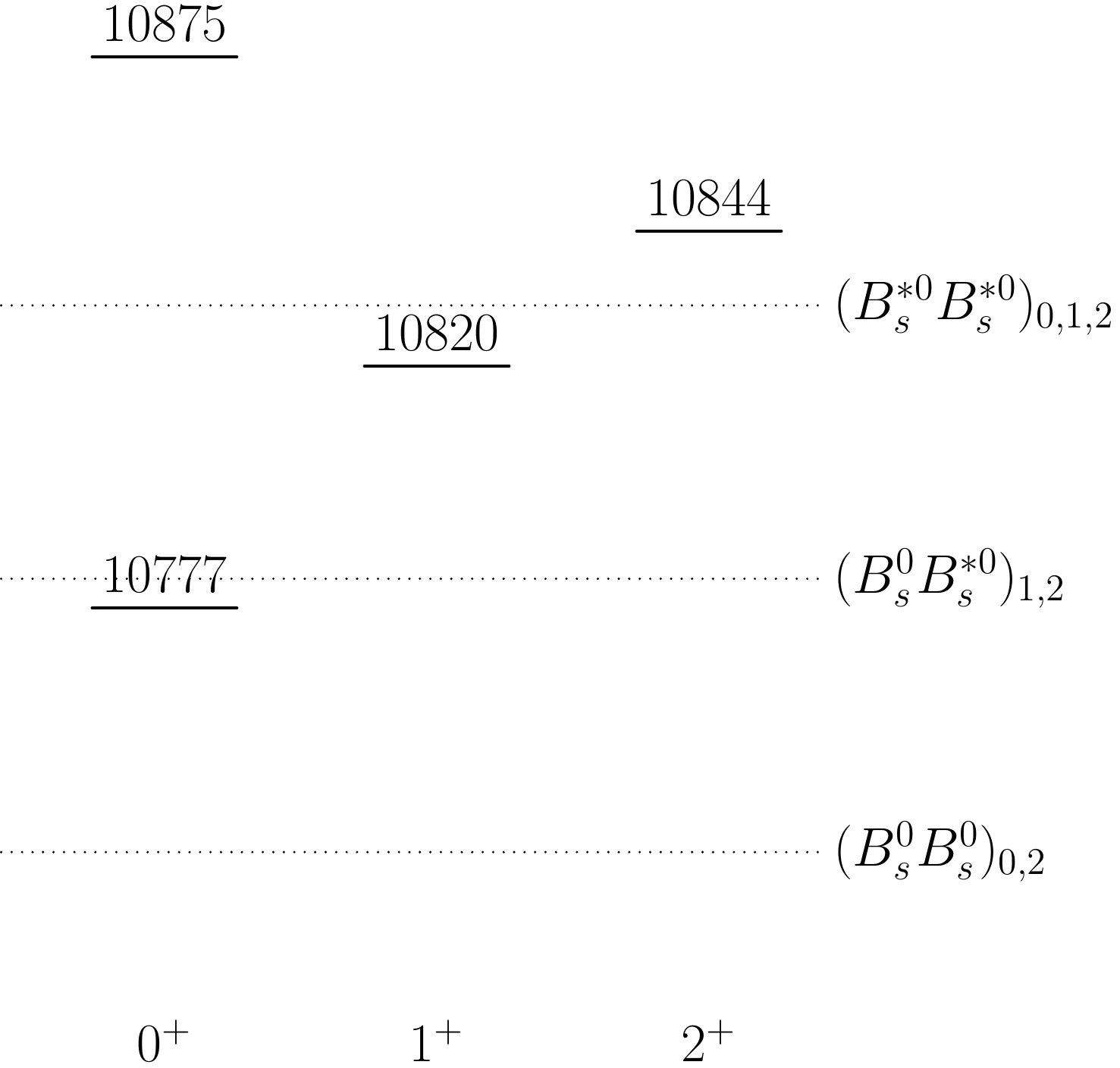}&\qquad&
\includegraphics[width=0.3\textwidth]{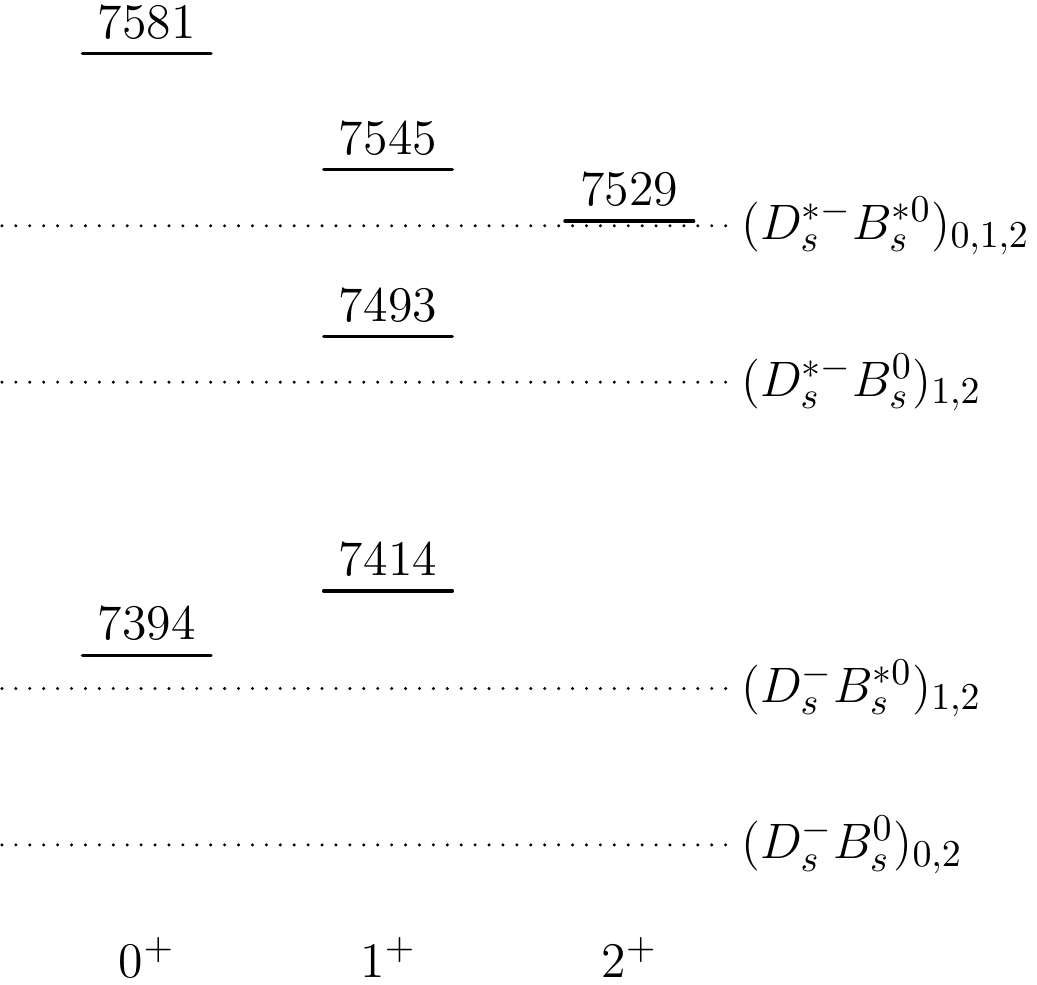}
\end{tabular}
\caption{The estimated masses (units: MeV) of the $ss\bar{c}\bar{c}$
(left), $ss\bar{b}\bar{b}$ (middle), and $ss\bar{c}\bar{b}$ (right)
systems in the threshold scheme. The dotted lines are thresholds of
the possible decay channels. When the total spin of a tetraquark is
equal to a subscript of the symbol for a meson-meson channel, the
corresponding $S$- or $D$-wave decay is allowed.}\label{figssQQ}
\end{figure*}

The expressions of the CMI matrices for the systems
$ss\bar{c}\bar{c}$ and $ss\bar{b}\bar{b}$ ($ss\bar{c}\bar{b}$) are
the same as those for the system $nn\bar{c}\bar{c}$
($nn\bar{c}\bar{b}$) with $I=1$, which has been shown in Tab.
\ref{class12} (\ref{class34}). As a result, the spectrum for
$ss\bar{Q}\bar{Q}$ has similar features to the isovector
$nn\bar{Q}\bar{Q}$ case. One may find the estimated masses and their
rough positions in Tab. \ref{tablessQQ} and Fig. \ref{figssQQ},
respectively. All the obtained tetraquarks except the tensor one
have S-wave decay channels and probably are broad states. The $2^+$
$ss\bar{c}\bar{c}$ ($ss\bar{c}\bar{b}$) is around the threshold of
$D_s^{*-}D_s^{*-}$ ($D_s^{*-}B_s^{*0}$). Its decay is mainly through
$D$ wave and may be a narrow state. The $2^+$ $ss\bar{b}\bar{b}$ is
slightly above the threshold of $B_s^{*0}B_s^{*0}$ and may have
narrower width compared with its $0^+$ and $1^+$ partner states.

\begin{table*}[htbp]\centering
\caption{Numerical results for the $ss\bar{c}\bar{c}$,
$ss\bar{b}\bar{b}$, and $ss\bar{c}\bar{b}$ systems in units of MeV.
The masses in the sixth column are estimated with the masses of
effective quarks and those in the last column with the thresholds of
$D_sD_s$/$B_sB_s$/$D_sB_s$.}\label{tablessQQ}

\begin{tabular}{c|c|c|c|c|c|c}\hline\hline
System&$J^P$&$\langle
H_{CM}\rangle$&Eigenvalues&Eigenvectors&Mass&$D_sD_s$/$B_sB_s$/$D_sB_s$\\\hline
$ss\bar{c}\bar{c}$&$0^+$&
$\left(\begin{array}{cc}-40.3&131.3\\131.3&46.8\end{array}\right)$&
$\left[\begin{array}{c}141.6\\-135.1\end{array}\right]$&
$\left[\begin{array}{c}(0.59,0.81)\\(-0.81,0.59)\end{array}\right]$&
$\left[\begin{array}{c}4672\\4395\end{array}\right]$&
$\left[\begin{array}{c}4293\\4016\end{array}\right]$\\

&$1^+$&-4.5&-4.5&1.00&4526&4146\\
&$2^+$&66.9&66.9&1.00&4597&4218\\\hline

$ss\bar{b}\bar{b}$&$0^+$&
$\left(\begin{array}{cc}0.3&45.1\\45.1&37.2\end{array}\right)$&
$\left[\begin{array}{c}67.4\\-30.0\end{array}\right]$&
$\left[\begin{array}{c}(0.56,0.83)\\(-0.83,0.56)\end{array}\right]$&
$\left[\begin{array}{c}11254\\11157\end{array}\right]$&
$\left[\begin{array}{c}10875\\10777\end{array}\right]$\\

&$1^+$&12.5&12.5&1.00&11199&10820\\
&$2^+$&37.1&37.1&1.00&11224&10844\\\hline


$ss\bar{c}\bar{b}$&$0^+$&
$\left(\begin{array}{cc}-22.1&88.2\\88.2&38.8\end{array}\right)$&
$\left[\begin{array}{c}101.6\\-85.0\end{array}\right]$&
$\left[\begin{array}{c}(0.58,0.81)\\(-0.81,0.58)\end{array}\right]$&
$\left[\begin{array}{c}7960\\7774\end{array}\right]$&
$\left[\begin{array}{c}7581\\7394\end{array}\right]$\\

&$1^+$&
$\left(\begin{array}{ccc}-9.3&16.6&-50.9\\16.6&1.9&35.2\\-50.9&35.2&21.2\end{array}\right)$&
$\left[\begin{array}{c}-65.5\\65.4\\13.8\end{array}\right]$&
$\left[\begin{array}{c}(0.67,-0.47,0.58)\\(-0.48,0.32,0.81)\\(0.57,0.82,0.01)\end{array}\right]$&
$\left[\begin{array}{c}7793\\7924\\7872\end{array}\right]$&
$\left[\begin{array}{c}7414\\7545\\7493\end{array}\right]$\\

&$2^+$&49.9&49.9&1.00&7908&7529\\\hline\hline
\end{tabular}
\end{table*}

\section{Discussions}\label{sec4}

\subsection{The spectrum}

We have explored various tetraquark configurations in a model with
the color-magnetic interaction, $QQ\bar{Q}\bar{Q}$ in
\cite{Wu:2016vtq}, $QQ\bar{Q}\bar{q}$ in \cite{Chen:2016ont}, and
$cs\bar{c}\bar{s}$ in \cite{Wu:2016gas}. A general feature for the
mass spectra of these systems is that both the lightest and the
heaviest states are those with $J^P=0^+$. This observation is also
valid for the $nn\bar{c}\bar{b}$, $ns\bar{c}\bar{b}$,
$ss\bar{c}\bar{c}$, $ss\bar{b}\bar{b}$, and $ss\bar{c}\bar{b}$
systems. In the $nn\bar{c}\bar{c}$, $nn\bar{b}\bar{b}$,
$ns\bar{c}\bar{c}$, and $ns\bar{b}\bar{b}$ cases, now the quantum
numbers of all the lightest states are $J^P=1^+$. Since the
meson-meson channels with the lowest threshold have the quantum
numbers $J^P=0^+$, the stable scalar tetraquarks must be below such
thresholds. Because of the angular momentum conservation, the
axial-vector tetraquarks above the lowest thresholds may also be
stable. This feature is unique for the $qq\bar{Q}\bar{Q}$ systems.

By comparing the diagonal elements of $\langle H_{CM}\rangle$ with
the eigenvalues, from Tables \ref{tablennQQ}, \ref{tablensQQ}, and
\ref{tablessQQ}, we may understand the color mixing effects for
various tetraquark states. The mass shifts due to the color mixing
range from several MeVs to a value around 100 MeV. The effects for
the $(nn\bar{c}\bar{c})^{I=1}$, $ns\bar{c}\bar{c}$, and
$ss\bar{c}\bar{c}$ states with $J^P=0^+$ are large while those for
the $(nn\bar{b}\bar{b})^{I=0}$ and $ns\bar{b}\bar{b}$ states with
$J^P=1^+$ are small. In a given system, the effects for the $0^+$
states are usually larger than the effects for the $1^+$ states.

For multiquark masses, the obtained values are just estimations
which rely on the parameters extracted from the conventional hadrons
and the adopted reference masses. As noted in Ref.
\cite{Park:2015nha}, some dynamical contributions are not
appropriately included in the present method and the tetraquark
masses need to be improved in a more elaborate model. One may guess
that not all the studied tetraquarks exist. However, if one state
could be observed, its partner states may be searched for with the
relative positions shown in Figs. \ref{fignnQQ}, \ref{fignsQQ}, and
\ref{figssQQ}. Such a study can be used to test the present model.

\subsection{Mass relations}

In principle, one may use the Gell-Mann--Okubo mass formula to
discuss the mass splittings in the same representation of flavor
$SU(3)$. In the present studied systems, the mixing between two
flavor representations occurs and it is difficult to find a general
mass formula for all the states. However, it is interesting that the
equal mass splitting relation
$m_{nn\bar{Q}\bar{Q}}+m_{ss\bar{Q}\bar{Q}}=2m_{ns\bar{Q}\bar{Q}}$ in
$6_f$ still holds for some states. In the $qq\bar{c}\bar{c}$ case,
the sets $[m_{nn\bar{c}\bar{c}}$, $m_{ns\bar{c}\bar{c}}$,
$m_{ss\bar{c}\bar{c}}]$=$[4128$, 4210, $4293]$ MeV and $[3850$,
3933, $4016]$ MeV for the $0^+$ states and the set
$[m_{nn\bar{c}\bar{c}}$, $m_{ns\bar{c}\bar{c}}$,
$m_{ss\bar{c}\bar{c}}]$=$[4044$, 4131, $4218]$ MeV for the $2^+$
states satisfy this relation. The set $[m_{nn\bar{c}\bar{c}}$,
$m_{ns\bar{c}\bar{c}}$, $m_{ss\bar{c}\bar{c}}]$=$[3973, 4060, 4146]$
MeV for the $1^+$ states also satisfies this relation. In the
$qq\bar{b}\bar{b}$ case, the sets satisfying this relation are:
$[m_{nn\bar{b}\bar{b}}$, $m_{ns\bar{b}\bar{b}}$,
$m_{ss\bar{b}\bar{b}}]$=$[10734$, 10804, $10875]$ MeV and $[10637$,
10707, $10777]$ MeV for the $0^+$ states, $[10694$, 10769, $10844]$
for the $2^+$ states, and $[10617$, 10718, $10820]$ MeV and
$[10671$, 10745, $10820]$ MeV for the $1^+$ states. In the
$qq\bar{c}\bar{b}$ case, the relation is roughly satisfied for the
sets: $[m_{nn\bar{c}\bar{b}}$, $m_{ns\bar{c}\bar{b}}$,
$m_{ss\bar{c}\bar{b}}]$=$[7428$, 7498, $7581]$ MeV and $[7241$,
7312, $7394]$ MeV for the $0^+$ states, $[7367$, 7442, $7529]$ MeV
for the $2^+$ states, and $[7393$, 7462, $7545]$ MeV, $[7332$, 7407,
$7493]$ MeV, and $[7258$, 7330, $7414]$ MeV for the $1^+$ states.

If one checks the mass formula in the estimation procedure, one
finds that the above mass relations actually reflect the flavor
SU(3) symmetry and its breaking. For example, for the $2^+$
$qq\bar{Q}\bar{Q}$ systems, the relation is the occasional result of
$C_{nn}+C_{ss}=2C_{ns}$ from which one may recover the relations
$2M_N+2M_{\Xi}=3M_\Lambda+M_\Sigma$ and
$M_{\Sigma^*}-M_\Delta=M_{\Xi^*}-M_{\Sigma^*}=M_{\Omega_c}-M_{\Xi^*}$.
Conversely, it is easy to understand the relation
$C_{nn}+C_{ss}=2C_{ns}$ in the CMI model from the flavor symmetry.
In principle, the equal mass relations in the mixing case should
also be from the relation $C_{nn}+C_{ss}=2C_{ns}$ and thus from the
flavor symmetry. As a byproduct of this relation, we get
$\Sigma^*-\Delta=\Xi-\Sigma$.

\subsection{Comparison with other works}

As mentioned in Sec .\ref{sec1}, there are also other works about
$qq\bar{Q}\bar{Q}$ tetraquark states. Various approaches such as the
non-relativistic quark model
\cite{SilvestreBrac:1993ss,Semay:1994ht}, the relativistic quark
model \cite{Ebert:2007rn}, and the QCD sum rule \cite{Du:2012wp}
have been used to explore the spectra. For comparison, we briefly
list the obtained results in the present model and those in other
theoretical methods in Table \ref{com}. One may consult Table V of
Ref. \cite{Ebert:2007rn} for more comparison. According to these two
tables, most masses are below the upper limits of the present
estimation (Scheme I) and are in the reasonable range. The upper
limits of masses for the $ss\bar{Q}\bar{Q}$ systems in QCD sum rule
seem to be high. Experimental investigations on tetraquark states
may be used to test theoretical methods according to their
predictions.

The isoscalar $T_{cc}$ is especially interesting. It is the most
promising tetraquark to be found first. Most works give a mass
around 3900 MeV. In the present work, we consider the configuration
mixing and get a lower mass around 3780 MeV (Scheme II). A
comparable value 3764 MeV was obtained in a chiral constituent quark
model in Ref. \cite{Vijande:2003ki}. Whether this $T_{cc}$ is stable
enough needs experimental judgement. Because of the limitations of
the present model (it is not a dynamical model, it has problems of
parameters, and so on), probably the mass is underestimated.

If all the predicted masses are 100 MeV underestimated, many
``stable'' states would be unstable. However, the lowest
$I(J^P)=0(0^+)$ $T_{cb}$ and the lowest $I(J^P)=0(1^+)$ $T_{bb}$ are
still below their respective thresholds of open-charm/bottom decay
channels, $B\bar{D}$ (7146 MeV) and $BB^*$ (10604 MeV), and seem to
be still stable. To search for these exotic tetraquarks in addition
to the $T_{cc}$ is also called for. Although the present model is
oversimplified, the basic features of spectra should be roughly
reasonable.

\begin{table*}
\caption{Comparison of results in various methods for tetraquark
states with the $qq\bar{Q}\bar{Q}$ configuration.}\label{com}
\begin{tabular}{cccccccc}\hline\hline
\multirow{2}{*}{System}&\multirow{2}{*}{$J$}&\multicolumn{4}{c}{Previous works}                         &\multicolumn{2}{c}{Our results}\\
                              &&\cite{SilvestreBrac:1993ss}&\cite{Semay:1994ht}&\cite{Du:2012wp,Chen:2013aba}&\cite{Ebert:2007rn}&Scheme 1&Scheme 2\\\hline

$(nn\bar{c}\bar{c})^{I=1}$
&0&--    &--               &--               &4056  &4078/4356       &3850/4128\\
&1&--    &--               &--               &4079  &4201             &3973\\
&2&--    &--               &--               &4118  &4271             &4044\\\\

$(nn\bar{c}\bar{c})^{I=0}$
&1&3931  &3892$\sim$3916   &--         &3935  &4007/4204   &3779/3977\\\\

$(nn\bar{b}\bar{b})^{I=1}$
&0&--    &--              &9900$\sim$10600  &10648  &10841/10937      &10637/10734\\
&1&10712 &--              &9900$\sim$10500  &10657  &10875&10671\\
&2&10735 &10710$\sim$10740 &--               &10673   &10897            &10694\\\\

$(nn\bar{b}\bar{b})^{I=0}$&1&10525 &10482$\sim$10514 &9900$\sim$10500  &10502      &10686/10821 &10483/10617\\\\

$(nn\bar{c}\bar{b})^{I=1}$
&0&--   &--               &7040$\sim$7380    &7383      &7457/7643       &7241/7428\\
&1&--   &--               &7030$\sim$7390    &7396/7403 &7473/7548/7609  &7258/7332/7393\\
&2&--   &--               &--                &7422      &7582            &7367\\\\

$(nn\bar{c}\bar{b})^{I=0}$
&0&7206 &7153$\sim$7183   &7040$\sim$7380    &7239      &7256/7429  &7041/7213\\
&1&7244 &7204$\sim$7231   &7030$\sim$7390    &7246      &7321/7431/7516  &7106/7215/7301\\
&2&7422 &--               &--                &--        &7530       &7315\\\\
$ns\bar{c}\bar{c}$
&0&--    &--               &3840$\sim$4300   &4221         &4236/4514        &3933/4210\\
&1&--    &--               &3950$\sim$5070   &4143/4239    &4225/4363/4400   &3921/4060/4096\\
&2&--    &--               &--               &4271         &4434             &4131\\\\

$ns\bar{b}\bar{b}$
&0&--    &--               &9900$\sim$10600  &10802        &10999/11095      &10707/10804\\
&1&10680 &10631$\sim$10665 &9900$\sim$11100  &10706/10809  &10911/11010/11037 &10619/10718/10745\\
&2&10816 &--               &--               &10823        &11060            &10769\\\\

$ns\bar{c}\bar{b}$
&0&--   &--               &--                &7444/7540    &7461/7615/7635/7801  &7158/7312/7332/7498\\
&1&--   &--               &--                &7451/7552/7555    &7530/7631/7634  &7227/7327/7330\\
& &     &                 &                  &                  &/7706/7710/7766 &/7402/7407/7462\\
&2&7496 &--               &--                &7572         &7718/7745       &7415/7442\\\\

$ss\bar{c}\bar{c}$
&0&--    &--               &4160$\sim$5200   &4359         &4395/4672        &4016/4293\\
&1&--    &--               &4010$\sim$5160   &4375         &4526             &4146\\
&2&--    &--               &--               &4402         &4597             &4218\\\\

$ss\bar{b}\bar{b}$
&0&--    &--               &9900$\sim$11200  &10932        &11157/11254      &10777/10875\\
&1&--    &--               &10100$\sim$11300 &10939        &11199            &10820\\
&2&--    &--               &--               &10950        &11224            &10844\\\\

$ss\bar{c}\bar{b}$
&0&--   &--               &7120$\sim$7450    &7673         &7774/7960       &7394/7581\\
&1&--   &--               &7210$\sim$7490    &7683/7684    &7793/7872/7924  &7414/7493/7545\\
&2&--   &--               &--                &7701         &7908
&7529\\\hline\hline
\end{tabular}
\end{table*}

\subsection{Production and decay}

The production of the $qq\bar{Q}\bar{Q}$ tetraquark states needs
high energy processes. For example, the C.M. energy for the
production of a $ss\bar{b}\bar{b}$ ($ss\bar{c}\bar{b}$) at an
electron-positron collider should be $\sqrt{s}>22$ ($\sqrt{s}>15$)
GeV. Of these exotic states, the production of the lowest $T_{cc}$
at various facilities (Tevatron, RHIC, LHC, KEK) has been considered
\cite{Vijande:2013qr,Hyodo:2012pm,hyodoetal,DelFabbro:2004ta,Cho:2010db,Cho:2011ew,Esposito:2013fma}.
Because of its clean background, the electron-positron collision
experiment has its advantage in searching for $T_{cc}$. At
Belle/BelleII, the $ns\bar{c}\bar{c}$ tetraquarks can also be
searched for.

From Figs. \ref{fignnQQ}, \ref{fignsQQ}, and \ref{figssQQ}, it is
easy to get a feature about the rearrangement decays and the
stability of the possible tetraquarks $qq\bar{Q}\bar{Q}$. For the
lowest stable states, the rearrangement decay channels are not
opened. One has to adopt weak or electromagnetic decay modes to search
for them: (1) for the lowest $T_{cc}^{I=0,J=1}$, one may use
$D^{*-}K^+\pi^-$
 or $D^-\bar{D}^0\gamma$ \cite{hyodoetal,Jin:2014nva}; (2) for the lowest $T_{bb}^{I=0,J=1}$, one may use $\bar{D}\pi B^{*}$, $\bar{D}^0D_s^+B^{*0}$, or $D^{*-}D_s^{*+}B^{*+}$; (3) for the lowest $T_{cb}^{I=0,J=0}$ ($T_{cb}^{I=0,J=1}$), one may use $\bar{D}\bar{D}\pi$ or $BK\pi$ ($\bar{D}\bar{D}^*\pi$ or $B^*K\pi$).; (4) for the lowest $(ns\bar{c}\bar{c})^{J=1}$, one may use $D_s^{*-}K\pi$, $\bar{D}^*K\bar{K}$, or $\bar{D}D_s^-\gamma$; (5) for the lowest $(ns\bar{b}\bar{b})^{J=1}$, one may use $\bar{D}\pi B_s^{*0}$, $B^*D_s^-\pi^+$, or $B^*D_s^{(*)+}D_s^{(*)-}$; (6) for the lowest $(ns\bar{c}\bar{b})^{J=0}$, one may use $K\pi B_s^0$, $BK\bar{K}$, $\bar{D}D_s^-\pi^+$, or $\bar{D}D_s^{(*)+}D_s^{(*)-}$; and (7) for the lowest $(ns\bar{c}\bar{b})^{J=1}$, one may use $K\pi B_s^{*0}$, $B^*K\bar{K}$, $\bar{D}\pi D_s^{*-}$, $\bar{D}^*D_s^-\pi^+$, or $\bar{D}^*D_s^{(*)+}D_s^{(*)-}$. For higher states, the two-body
meson-meson channels are the dominant decay modes. Searching for
them in meson-meson channels can probably give interesting exotic
signals. In addition, their transition to lower tetraquarks is also
allowed, e.g. an $I(J^P)=1(1^+)$ $nn\bar{c}\bar{b}$ state may decay
to an $I(J^P)=0(0^+)$ $nn\bar{c}\bar{b}$ state by emitting a pion.
However, the decay into OZI-allowed
(light-baryon)+(heavy-antibaryon) final state is forbidden by
kinematics even if the doubly-heavy baryons are observed.

\subsection{``Good'' diquark and stable tetraquarks}

The studies on various multiquark states can be found in the literature. Usually, the multi-body problem is simplified by assuming the existence of possible substructures, e.g. diquark or triquark. In the conventional baryons, the color-antitriplet diquark with $spin=0$ is called ``good'' diquark while that with $spin=1$ is called ``bad'' one because the color-spin interaction in the former (latter) case is attractive (repulsive). An example because of their difference is the mass splitting between $\Sigma_c$ and $\Lambda_c$. In the multiquark study, one expects that states containing the scalar diquark have a lower mass and are easy to search for experimentally while those with the axial-vector diquark should be broad resonances even if they exist and are difficult to search for. 

Because the interaction related with heavy quarks is suppressed, the properties of the tetraquarks $qq\bar{Q}\bar{Q}$ are determined mainly by the light diquark. In the present work, the number of states is significantly reduced if we consider tetraquarks containing the light ``good'' diquark only. Now the retained color-spin bases in Eq. (\ref{basis-vectors}) are $\phi_2\chi_5$ (total spin=1) and $\phi_2\chi_6$ (total spin=0). From tables \ref{class12}-\ref{class56}, there is no ``good'' light diquark in the $ss\bar{Q}\bar{Q}$ tetraquark states and one needs to discuss only eight lowest states: $1^+$ $nn\bar{c}\bar{c}$, $1^+$ $nn\bar{b}\bar{b}$, $0^+$ and $1^+$ $nn\bar{c}\bar{b}$, $1^+$ $ns\bar{c}\bar{c}$, $1^+$ $ns\bar{b}\bar{b}$, and $0^+$ and $1^+$ $ns\bar{c}\bar{b}$. Without channel coupling, their masses are higher than the lowest states in Figs. \ref{fignnQQ} and \ref{fignsQQ}. We list their values in table \ref{gooddiquark} with the threshold scheme. Now, the $1^+$ $nn\bar{c}\bar{b}$ state and the two $ns\bar{c}\bar{b}$ states are slightly above their corresponding thresholds of rearrangement decay channels. Stable states with the configurations $nn\bar{c}\bar{c}$, $nn\bar{b}\bar{b}$, $nn\bar{c}\bar{b}$, $ns\bar{c}\bar{c}$, and $ns\bar{b}\bar{b}$ are still possible.
\begin{table}
\caption{Numerical results (units: MeV) for the systems containing light ``good'' diquark only. The tetraquark masses in the last column are estimated with the thresholds of $DD/BB/BD/DD_s/BB_s/DB_s$.}\label{gooddiquark}
\begin{tabular}{cccc}\hline\hline
System & $J^P$ & $\langle H_{CM}\rangle$ & $DD/BB/BD$\\
&&&$/DD_s/BB_s/DB_s$\\\hline
$(nn\bar{c}\bar{c})^{I=0}$&$1^+$& -132.3&3813\\
$(nn\bar{b}\bar{b})^{I=0}$&$1^+$& -138.7&10487\\
$(nn\bar{c}\bar{b})^{I=0}$&$0^+$& -172.8&7113\\
$(nn\bar{c}\bar{b})^{I=0}$&$1^+$& -137.6&7149\\
$(ns\bar{c}\bar{c})$&$1^+$&-84.3&3964\\
$(ns\bar{b}\bar{b})$&$1^+$&-84.3&10626\\
$(ns\bar{c}\bar{b})$&$0^+$&-124.8&7252\\
$(ns\bar{c}\bar{b})$&$1^+$&-89.6&7287\\\hline\hline
\end{tabular}
\end{table}

However, as shown in Ref. \cite{Hyodo:2012pm}, the color-sextet diquark with $spin=1$ also has weak attraction but it does not exist in the conventional baryons. The contributions from such diquark and other diquarks other than the ``good'' diquark lead to more complex mass spectra of possible multiquark states. In the present study, we use ``diquark'' to denote the two-quark cluster which may have various quantum numbers. Our analysis shows that the channel coupling between the $6_c\otimes\bar{6}_c$ and $\bar{3}_c\otimes3_c$ color configurations further lowers the $qq\bar{Q}\bar{Q}$ state where the $qq$ pair forms a scalar diquark.

\section{Summary}\label{sec5}

In this work, we systematically study the mass spectra of tetraquark
states with the $qq\bar{Q}\bar{Q}$ configuration by using a simple
color-magnetic model. All possible quantum numbers without orbital
excitation are considered. We find that the color mixing effects are
relatively important for the $0^+$ states. The effect for the lowest
$T_{cc}$ ($T_{bb}$) is about 30 (7) MeV and is not large. If the
results shown in Figs. \ref{fignnQQ}, \ref{fignsQQ}, and
\ref{figssQQ} are all reasonable, two types of $qq\bar{Q}\bar{Q}$
tetraquarks are probably stable: (1) $T_{cc}^{I=0}$, $T_{bb}^{I=0}$,
$T_{cb}^{I=0}$, $ns\bar{c}\bar{c}$, $ns\bar{b}\bar{b}$, and
$ns\bar{c}\bar{b}$ with $J^P=1^+$; and (2) $T_{cb}^{I=0}$ and
$ns\bar{c}\bar{b}$ with $J^P=0^+$. The feature that the stable
states could have $J=1$ is unique compared with other tetraquark
structures $QQ\bar{Q}\bar{Q}$, $QQ\bar{Q}\bar{q}$, and
$cs\bar{c}\bar{s}$. Based on our estimations, possible strong decay
patterns are briefly discussed. Up to now, none of the exotic
$qq\bar{Q}\bar{Q}$ states has been observed and more detailed
investigations on their properties are still needed. Hopefully, our
study on these interesting tetraquark states may be helpful to the
future experimental searches.

\begin{acknowledgements}
This paper is supported by the National Natural Science Foundation
of China under Grant No. 11175073, No. 11275115, No. 11222547, No.
11261130311 and 973 program. Xiang Liu is also supported by the
National Program for Support of Top-notch Young Professionals and
the Fundamental Research Funds for the Central Universities.
\end{acknowledgements}



\end{document}